\documentclass[%
 reprint,
 amsmath,amssymb,
 aps,
longbibliography
]{revtex4-2}

\usepackage{graphicx}% Include figure files
\usepackage{dcolumn}% Align table columns on decimal point
\usepackage{bm}% bold math
\usepackage{blkarray}

\usepackage{xcolor}

\usepackage{hyperref}       % hyperlinks

\begin{document}
\title{Discovering hidden layers in quantum graphs}

\author{\L{}ukasz G. Gajewski}
% \email{lukaszgajewski@tuta.io}
\affiliation{Faculty of Physics, Warsaw University of Technology, Koszykowa 75, 00-662 Warszawa, Poland}
 \author{Julian Sienkiewicz}
 \affiliation{Faculty of Physics, Warsaw University of Technology, Koszykowa 75, 00-662 Warszawa, Poland}
 \author{Janusz A. Ho\l{}yst}
 \affiliation{Faculty of Physics, Warsaw University of Technology, Koszykowa 75, 00-662 Warszawa, Poland}
 \affiliation{ITMO University, Kronverkskiy Prospekt 49, St Petersburg, Russia 197101}

\begin{abstract}
Finding hidden layers in complex networks is an important and a non-trivial problem in modern science. We explore the framework of quantum graphs to determine whether concealed parts of a multi-layer system exist and if so then what is their extent, i.e., how many unknown layers are there. 
Assuming that the only information available is the time evolution of a wave propagation on a single layer of a network it is indeed possible to uncover that which is hidden by merely observing the dynamics. 
We present evidence on both synthetic and real-world networks that the frequency spectrum of the wave dynamics can express distinct features in the form of additional frequency peaks. 
These peaks exhibit dependence on the number of layers taking part in the propagation and thus allowing for the extraction of said number. 
We show that in fact, with sufficient observation time, one can fully reconstruct the row-normalised adjacency matrix spectrum. 
We compare our propositions to a machine learning approach using a wave packet signature method modified for the purposes of multi-layer systems.

%We compare this approach to the work of Aziz \textit{et al.} in which there has been established a wave packet signature method for discerning between various single layer graphs and we modify it for the purposes of multi-layer systems.
\end{abstract}

\maketitle
% keywords can be removed
% \keywords{First keyword \and Second keyword \and More}

\section{Introduction}
A plethora of contemporary dynamical systems and collective phenomena can be expressed in terms of complex networks and recently often with multi-layer models. Whether it is a transportation network (e.g., buses and trams) or a social one (e.g., Twitter and Facebook), various forms of information propagation or state dynamics can be described with multi-layer networks. \cite{de2013mathematical, de2016physics, de2018fundamentals, kivela2014multilayer,boccaletti2014structure, barrat2008dynamical, pastor2015epidemic}

However, it is not uncommon for certain parts of a system to remain hidden from observers, and it can be crucial to be able to discern what characteristics are unknown and then to try to obtain them. Such inverse problems have been studied in various settings in both topology and dynamics parameters reconstruction in mono- and multi-layer scenarios alike \cite{lokhov2016reconstructing, wilinski2020scalable,leskovec2010inferring, woo2020iterative, pouget2015inferring, abrahao2013trace, gripon2013reconstructing, gomez2012inferring, netrapalli2012learning, braunstein2019network,  gajewski2019multiple, paluch2018fast, paluch2020optimizing}.
Recently there have been some pivotal advances in terms of determining whether one can detect hidden layers in non-markovian dynamics and even obtain how many of these layers are there \cite{Lacasa_2018}. In the same manner hidden layers can be detected in the case of epidemic like processes (e.g., SI --- susceptible-infected or IC -- independent-cascades)  \cite{Gajewski2021pre}.

Nevertheless, to our knowledge, there has not been much done specifically for uncovering the multi-layer structures in quantum graphs and thus we address this issue. We present two potentially viable approaches of establishing if hidden layers exist, and in some scenarios to ascertain the exact count of these layers. 

Traditionally, graphs are discrete, combinatorial abstract mathematical objects. If we supply them with a metric and topology we call such objects \textit{metric graphs}. Those in turn equipped with a  second order differential operator acting on its vertices and edges - a Hamiltonian - and appropriate boundary conditions are called \textit{quantum graphs} \cite{kuchment_quantum_2004, kuchment_quantum_2005, kuchment_quantum_2008, berkolaiko_introduction_2013}.
It is worth underlining here that with this definition we do not specify the exact nature of the Hamiltonian and while it is often a quantum mechanical one it need not be so and thus here we follow the interpretation of taut strings, fused together at the vertices that can be seen as the ``limiting case'' of a ``quantum wire'' \cite{friedman_wave_2004, hurt2013mathematical}.
The (most likely) first use of this framework can be traced back to Pauling's paper in 1936 \cite{pauling1936diamagnetic}, however, for the most part quantum graphs have not been widely used until more recently. Nowadays they see many various applications in dynamical systems, nanotechnology, photonic crystals and many others \cite{kuchment2002graph, biamonte_complex_2019, exner2008quantum, faccin_community_2014, faccin_degree_2013, cuquet_entanglement_2009}.
Most recently Aziz \textit{et al.} established a method based on a wave packet propagation on quantum graphs that allows to distinguish between structures in complex networks \cite{aziz_wave_2019} thanks to many well studied properties of the Laplacian (e.g., a finite speed of propagation \cite{kostrykin_finite_2012} as opposed to a discrete Laplacian \cite{gomez2013diffusion, sole2013spectral}) and its spectra in quantum graphs \cite{aziz2013analysis, aziz2013graph, WILSON20134183, pesenson_analysis_2006, pesenson_band_2005, exner_spectral_2013, cattaneo_spectrum_1997}. The idea of determining the shape of an object based on observable dynamics on it goes back to the work of Kac in 1966 \cite{kac1966can} in which he asks whether it is possible to hear the shape of a drum. Giraud and Thas showed that the eigenvalues of different shapes can be identical and therefore answered Kac's question in the negative. Gutkin and Smilansky, on the other hand, showed that in quantum graphs specifically, under certain conditions, one can indeed ``hear'' the shape as the Hamiltonian uniquely defines the connections and their lengths when the graph is finite (and simple), the bond lengths are rationally independent and the vertex scattering matrices are properly connecting. 
It is also worth noting that this inverse spectral problem can be extended onto scattering systems as also stated in the same paper (a so called inverse scattering problem \cite{gordon2005isospectral}).
However, in these scattering systems it appears that it is \textit{not} always possible to uniquely find the structure \cite{band2010scattering, band2011note}, i.e., there is a way to construct iso-scattering pairs of graphs with identical polar structure of their scattering matrices, which was also shown experimentally via microwave networks by Hul \textit{et al.} \cite{Hul2012are}. 
Wave packets specifically have also attracted some attention in recent years but not for the purposes of what we aim for in this paper \cite{smilansky2017delay, smilansky2018delay}.
Some work has been done in the context of sufficient coverage with sensors \cite{robinson2010inverse}, however, in this case we will not share all the assumptions and thus those methods are not applicable to our problem.

 \begin{figure}[tb]
     \includegraphics[width=.35\textwidth]{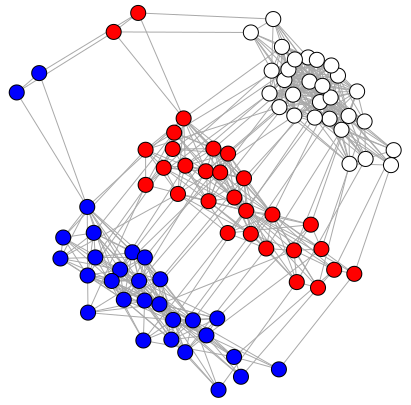}
     \caption{Three-layered multiplex representation of the Vickers \cite{vickers1981representing} data.}
     \label{fig:multiplex}
 \end{figure}
 
In this paper we tackle the problem of determining whether there are hidden layers in the complex system we are observing and if so then how many. We assume wave packet propagation dynamics on a quantum graph as our model for the dynamical system.
Each edge $e$ in the graph $G$ has an associated length $l_e=1$ and a spatial coordinate variable $x_e \in [0, l_e]$ along said edge. We use a special case of a Hamiltonian - an edge based Laplacian giving us an edge-based wave equation on a graph in the form:
\begin{equation}
    u_{tt}d\mathcal{E} = -\Delta u
\end{equation}
where $\mathcal{E}$ is a Lebesgue measure on the graph's edges \cite{friedman_wave_2004, friedman_calculus_2004}, and $u$ is a square integrable function defined on the graph. Specifically $u(n)$ is the value of $u$ at the node $n$, and $u(e, x_e)$ is the value of $u$ at the position $x_e$ along the edge $e$ \cite{aziz_wave_2019}.

We use the Neumann boundary conditions stating that the sum of the outward pointing gradients at every vertex must vanish \cite{friedman_wave_2004, aziz_wave_2019}:
\begin{equation}
    \forall v \in G, ~\sum_{e \ni v} (-1)^{1-x_{e,v}} \nabla u(e, x_{e,v}) = 0
\end{equation}
The initial condition for the wave equation is a Gaussian wave packet:
\begin{equation}
    u(e, x) = \exp{\big(-a(x-\mu)^2\big)}
\end{equation}
which is fully contained within a single edge with the highest betweenness centrality, i.e., such an edge that goes through as many shortest paths in the graph as possible \cite{brandes2001faster, newman2005measure}, following the conventions of Aziz \textit{et al.} \cite{aziz_wave_2019}. 
We simulate the many-layer system in a multiplex configuration, i.e. we consider a system with a set of $L \geq 1$ interacting layers (mono-layer networks), where each of them has the same set of $N$ nodes but different topologies (set of connections), and each node is connected to its reflection (also sometimes called a replica) in a neighbouring layer
(see Fig.~\ref{fig:multiplex} for a real-world network example and Fig.~\ref{fig:multi_example} for a wave propagation example on a simple synthetic graph). In its simplest form a multi-layer system can be represented with a set of intra-layer adjacency matrices $\{\mathbf{C}^l\}$ ($\forall_l \mathbf{C}^l \in \mathcal{R}^{N\times N}$), where $\mathbf{C}^l_{i,j} = 1$ when nodes $i, j$ are connected in the layer $l$ and $0$ otherwise. Additionally we need a set of inter-layer adjacency matrices $\{\mathbf{B}\}$ ($\mathbf{B} \in \mathcal{R}^{L\times L}$), where the elements $\mathbf{B}_{l_1, l_2}$ signify whether the layers $l_1,l_2$ are connected (i.e., each node from layer $l_1$ is connected to its replica in layer $l_2$).
$\{\mathbf{C}^l\}$ and $\{\mathbf{B}\}$ can then be combined into a single adjacency matrix $\mathbf{A} \in \mathcal{R}^{LN\times LN}$. For an example of how that can be done see Appendix \ref{app:full}, and for a more elaborate tensor formulation (unnecessary for our purposes here) see Ref.\cite{de2013mathematical}.
While the propagation simulations are computed on full systems, for the detection purposes we always only use information from a single layer, i.e., all but one layer are hidden from the perspective of our methods at all times.

Although the idea of travelling waves in the network structure might seem to be academic and considered artificial as being far from typically considered dynamics such as epidemic \cite{Paluch2021} or opinion spreading \cite{Chmiel2017,Chmiel2020}, it is essential to note that the wave can simply model the information propagating due to interdependences between these concepts \cite{Franceschetti2017}. Also in the case of networks, certain propagation dynamics connected to shock waves \cite{Mones2015} or excitable nodes \cite{Liao2011,Liao2011a} have been successfully undertaken. Therefore our study brings very specific applications for a variety of real-world systems.

The rest of this paper consists of three main parts followed by a discussion. Firstly, we briefly outline the approach introduced by Aziz \textit{et al.} in \cite{aziz_wave_2019} and show its viability for the purposes of multi-layer networks in the context described above. Secondly, we introduce our own approach with the use of a Fourier transform on the time evolution of the sum of the amplitudes in the visible part of the system. Thirdly, we show that with sufficient resources (i.e., observation time) one can fully reconstruct the spectrum of the row-normalised adjacency matrix.

\section{Gaussian wave packet signature (WPS)}
\begin{figure*}[tb]
     \includegraphics[width=0.45\textwidth]{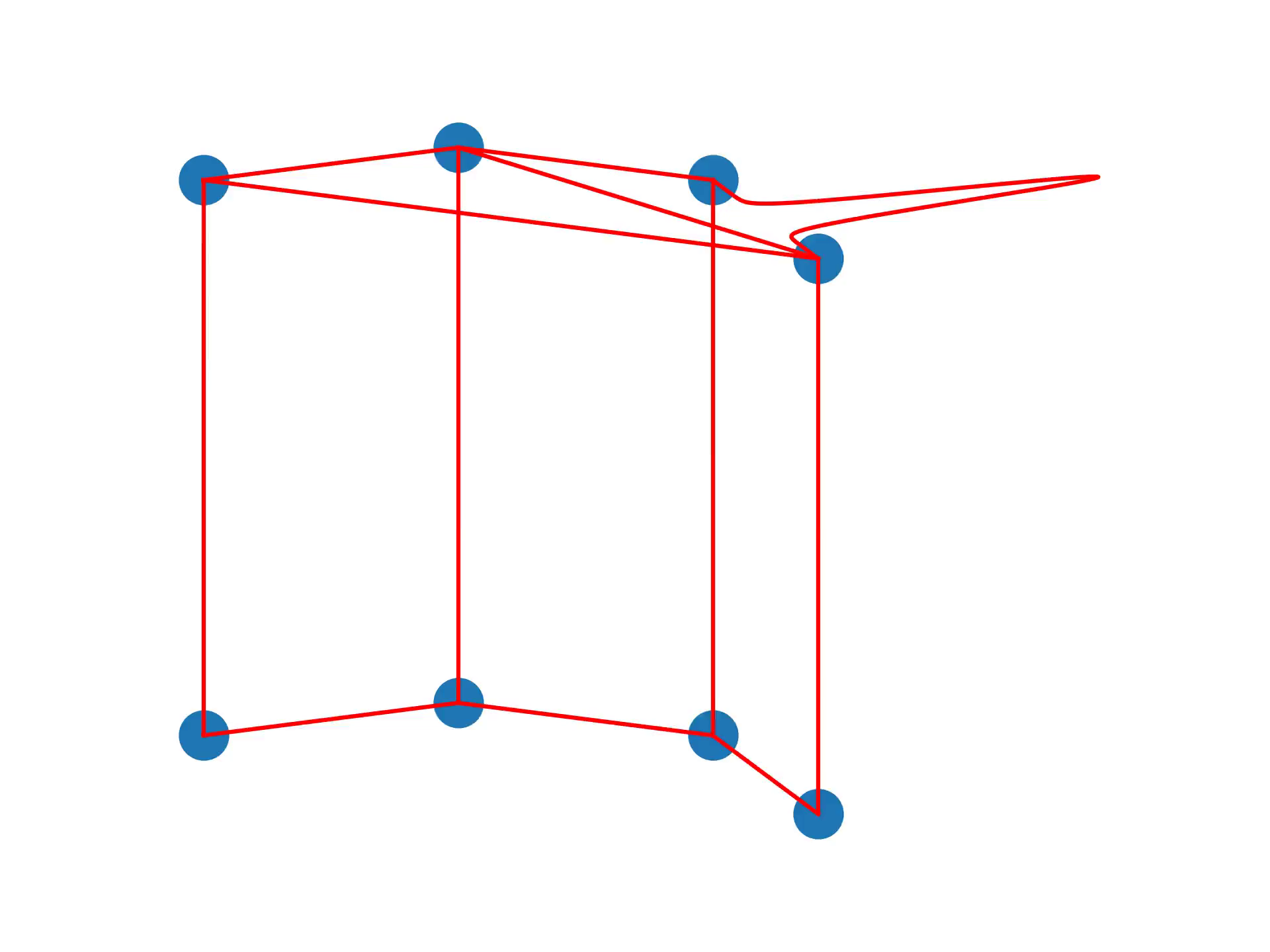}
     \hfill
     \includegraphics[width=0.45\textwidth]{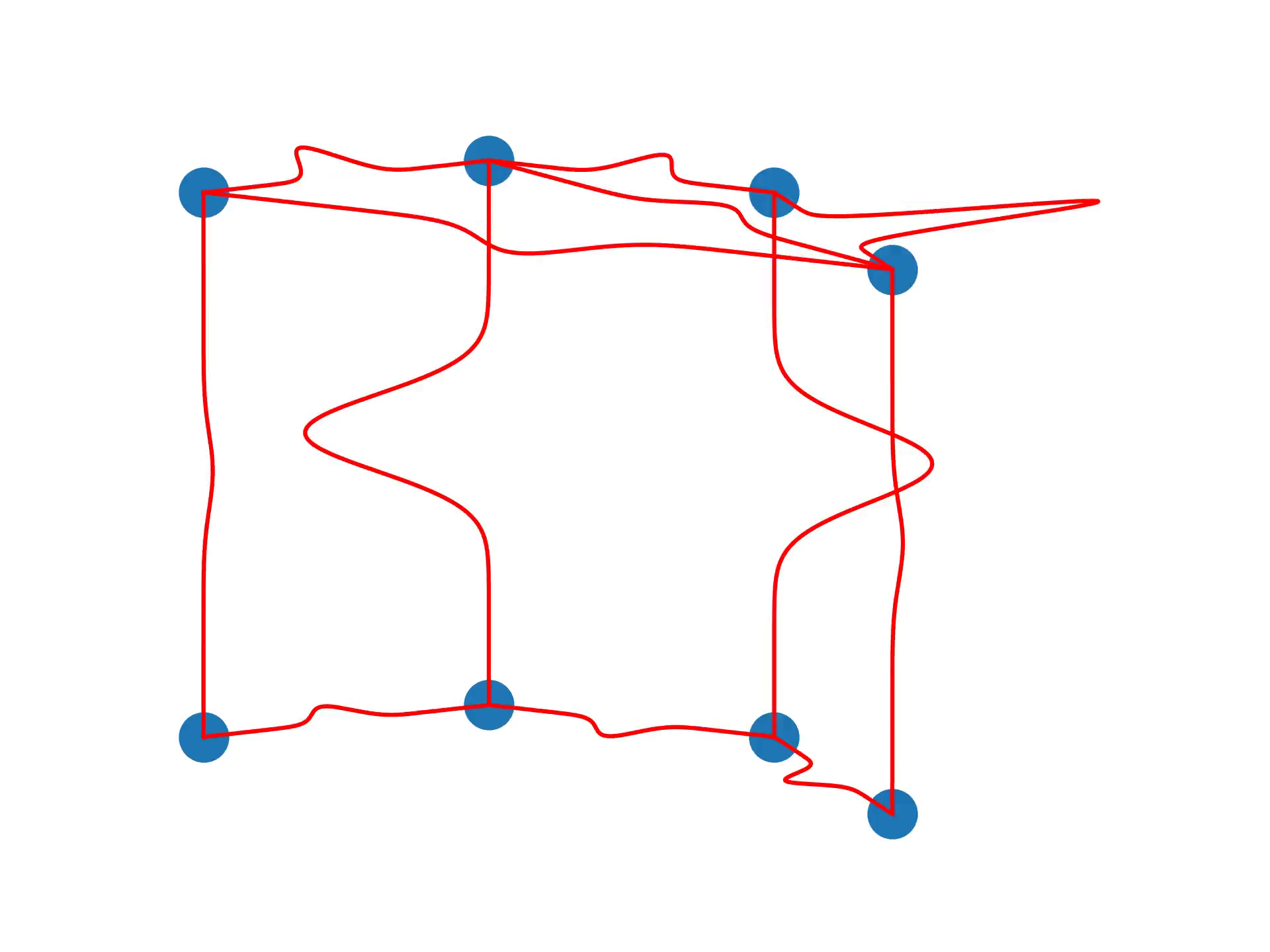}
    \caption{Illustration of a wave starting at a specific edge (left, $t=0$) and then propagating through the multiplex system (right, $t>0$).}
    \label{fig:multi_example}
\end{figure*}

Gaussian wave packet signature (WPS) is a methodology developed by Aziz \textit{et al.} \cite{aziz_wave_2019} that allows to distinguish between various types of graphs. 
The procedure starts by initiating an edge with a Gaussian wave packet that is completely contained on said edge (see the left panel in Fig. \ref{fig:multi_example}). The edge is chosen to be the one with the highest betweenness centrality to assure the fastest possible spread of the wave on the graph, although this can, of course, be relaxed in general. The choice of the edge from which the propagation starts does not impact either of the presented methods performance (including our own Fourier transform based described later).
All this does is simply speeding up our testing set up, while in practice it would even be possible to apply the methods without knowing (or choosing) the primary edge at all. Additionally this is the choice presented in \cite{aziz_wave_2019}, and we follow that for the sake of continuity and ease of comparison. 
Then, on each integer time we measure the amplitude in the centre of every edge $2|E|$ times in total, where $|E|$ is the number of edges in the known layer (see the right panel in Fig. \ref{fig:multi_example} and Appendix \ref{app:a} for a detailed description of the way the amplitude is calculated). 
Again, the number of measurements has been selected for the sake of comparison with \cite{aziz_wave_2019} where such a value had been used. The difference here is that Aziz et al. assume knowledge of the whole graph and $|E|$ there refers to all edges in the system, whilst for us it refers only to the visible layer. 
We measure the centre of each edge because at integer times the highest value is in the centre. Finally, we create a histogram with $100$ bins of these measurements - this is the WPS of the graph. 
Aziz \textit{et al.} show that particular graph types (such as the evolving preferential attachment graph model of Barab{\'a}si and Albert (BA) \cite{barabasi1999emergence} or the static random graph model of Erd{\H{o}}s and R{\'e}nyi (ER) \cite{erdHos1960evolution}, etc.) will have similar WPSs yet different in comparison to other types (so, e.g., one can differentiate an ER from a BA). In order to actually do this differentiation one must build a classifier. In their work, a K-nearest neighbours (K-NN) classifier was chosen. From the perspective of the machine learning tools we use in this paper (K-NN, PCA) each histogram bin of the WPS is a dimension in the feature space.

For our purposes, we will deviate slightly from this procedure. 
Namely, to us the whole graph is not known and the graph itself is a multiplex. Additionally, we assume that the wave propagation is an actual process ongoing through some network. Thus, we assume we have access to a single layer on which a certain spread has happened that can be modelled with a Gaussian wave packet and we suspect there may be hidden layers in the network. The question is - can we detect their presence?

\begin{figure*}[tb]
     \includegraphics[width=0.3\textwidth]{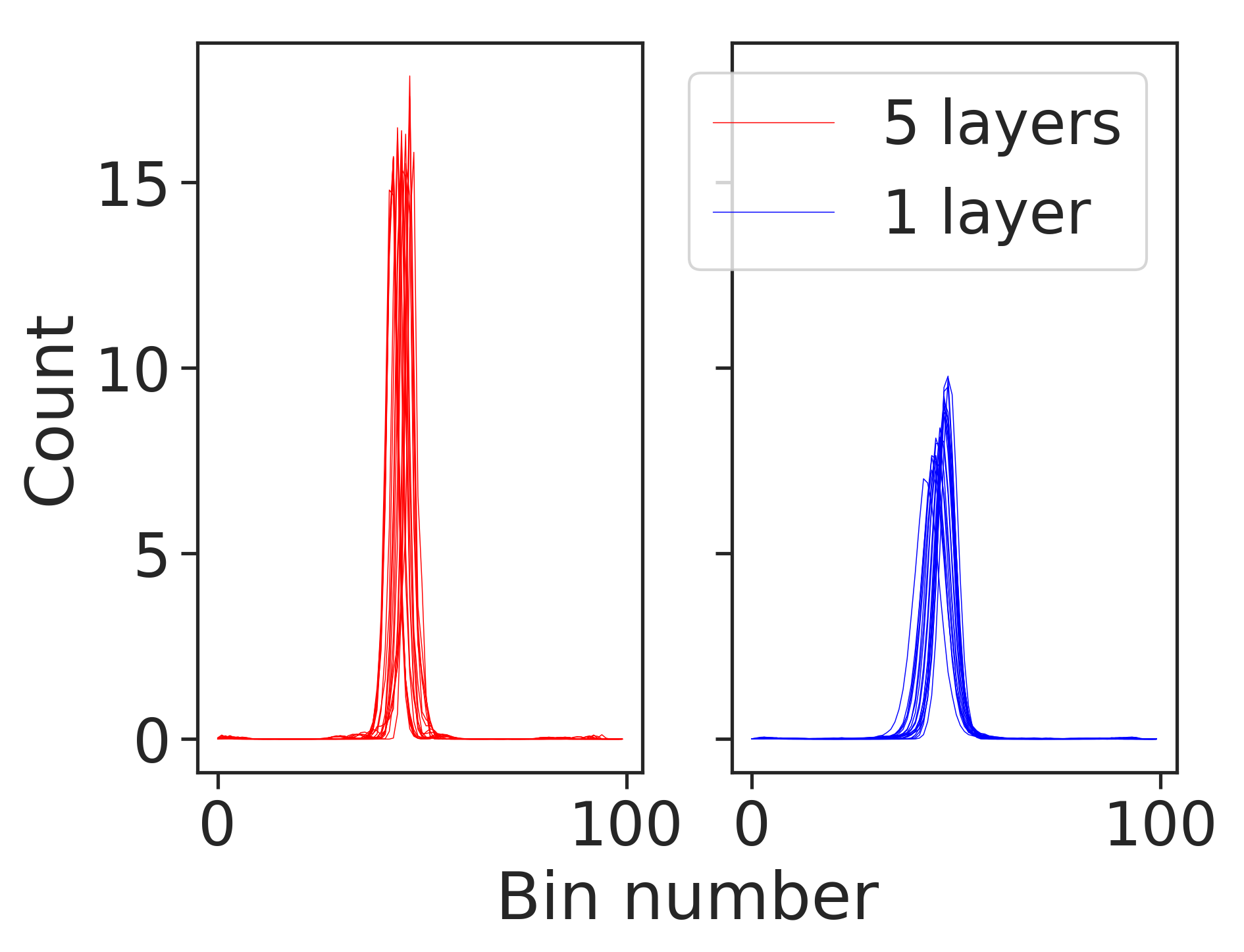}
     \hfill
     \includegraphics[width=0.3\textwidth]{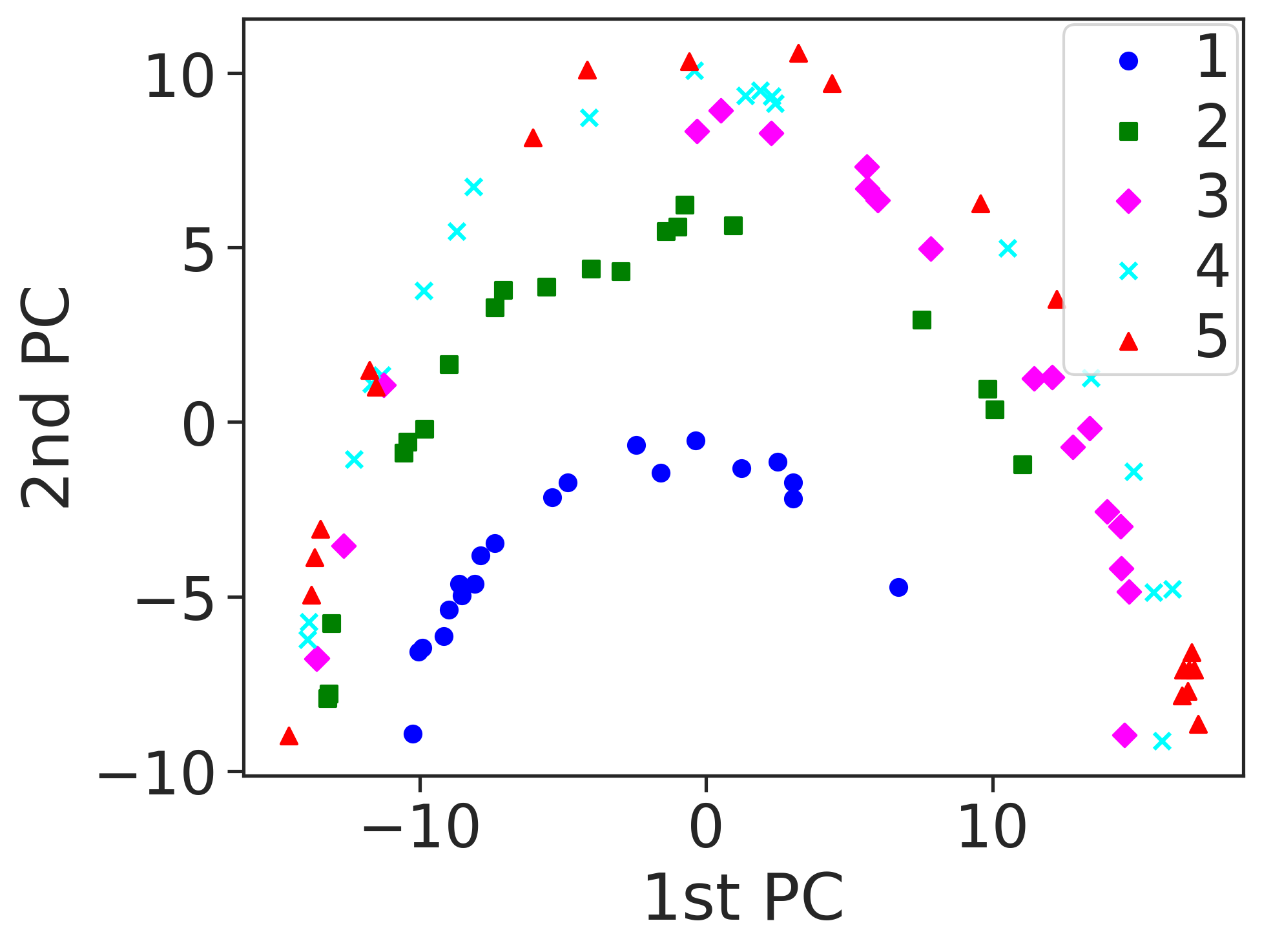}
     \hfill
     \includegraphics[width=0.3\textwidth]{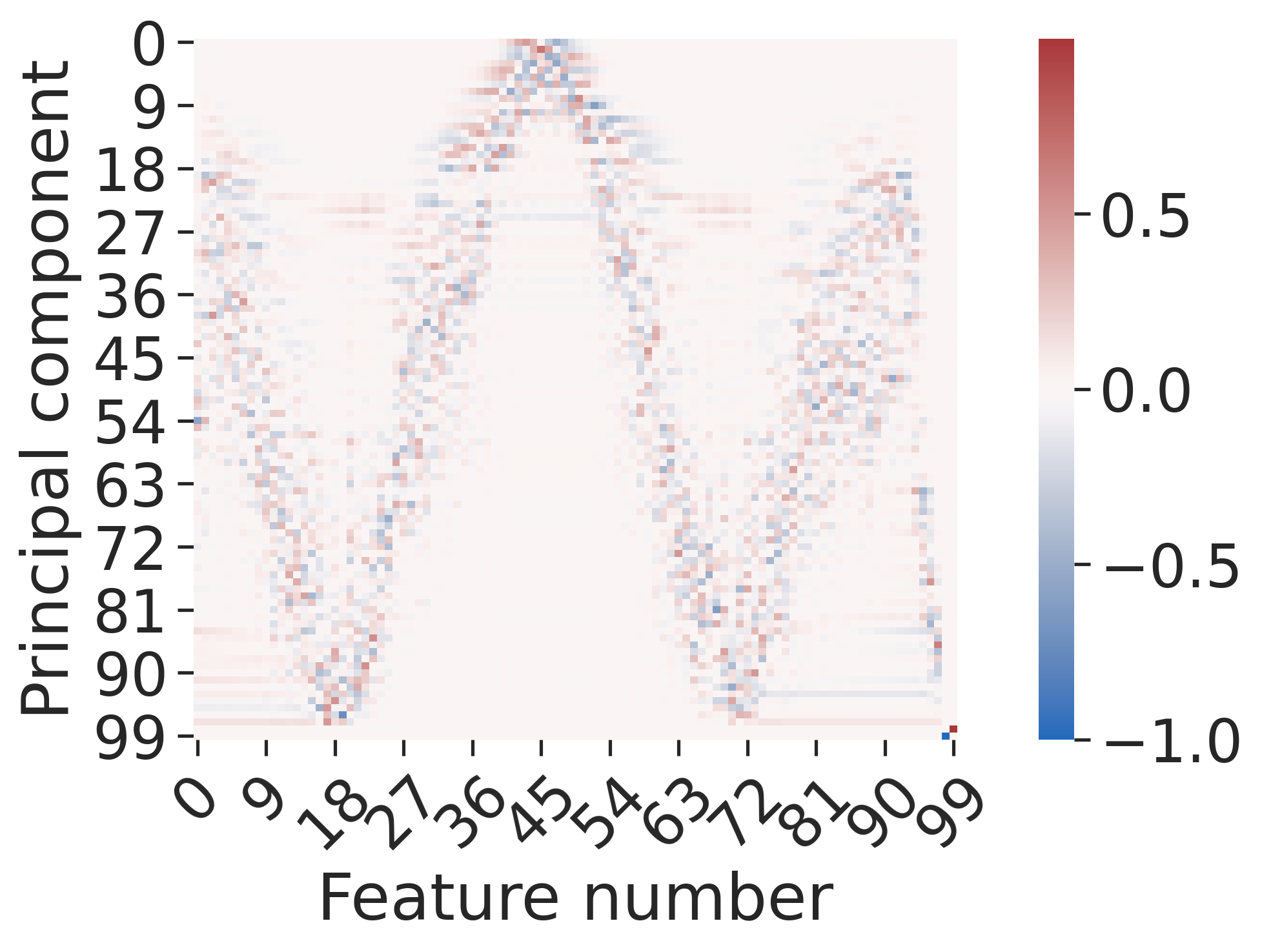}
    \caption{Wave packet signatures (left) for various realisations of a Barab{\'a}si-Albert ($N=50$, $m=3$) graph with 5 and 1 layers (measured on one layer only). WPS projected onto a 2D space with PCA (centre) where colours and symbols distinguish the no. of layers. Each point is a different BA graph. PCA transformation matrix (right) showing the contributions of given WPS bins into principal components.}
    \label{fig:wps-ba}
\end{figure*}

\begin{figure*}[tb]
    \includegraphics[width=0.3\textwidth]{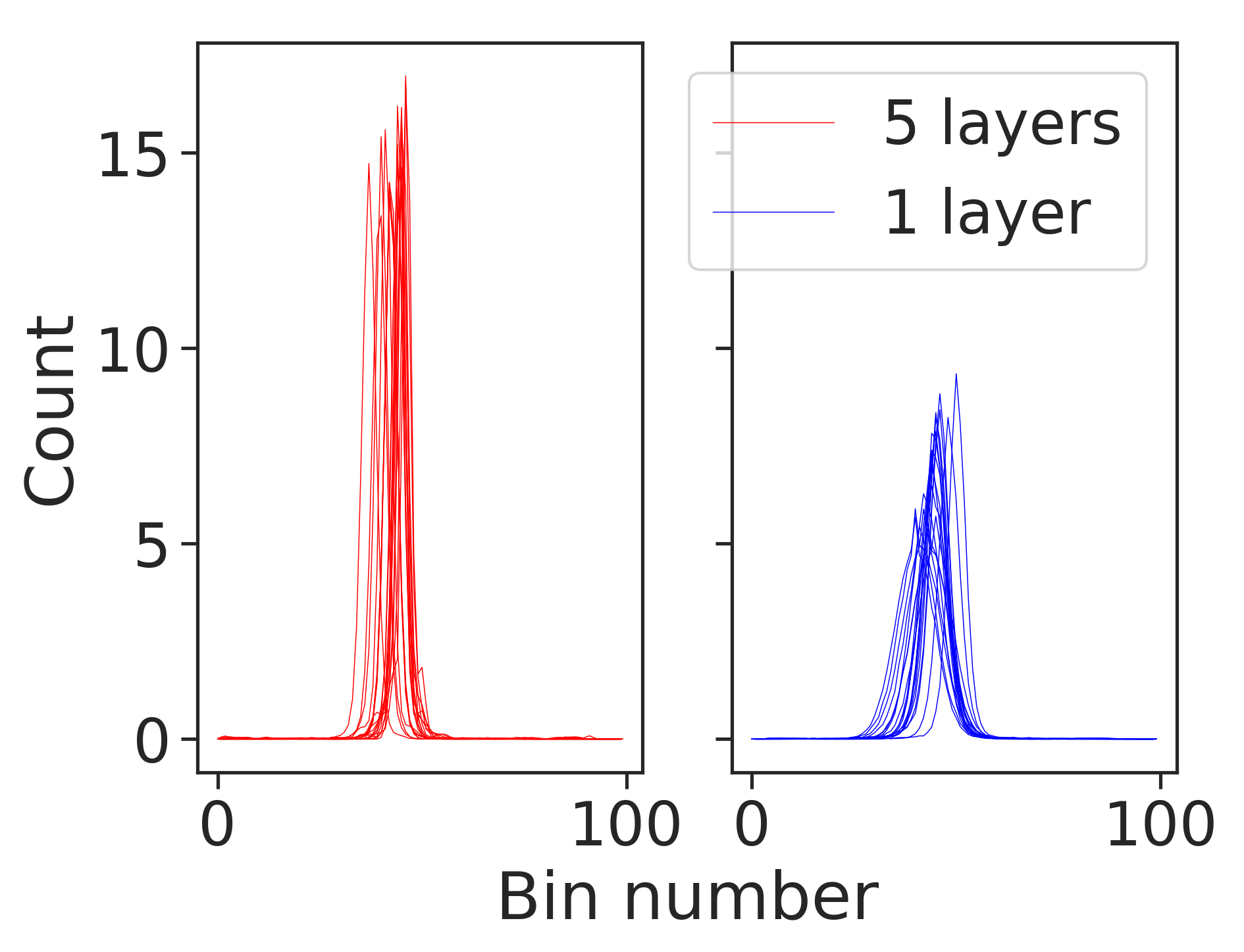}
    \hfill
    \includegraphics[width=0.3\textwidth]{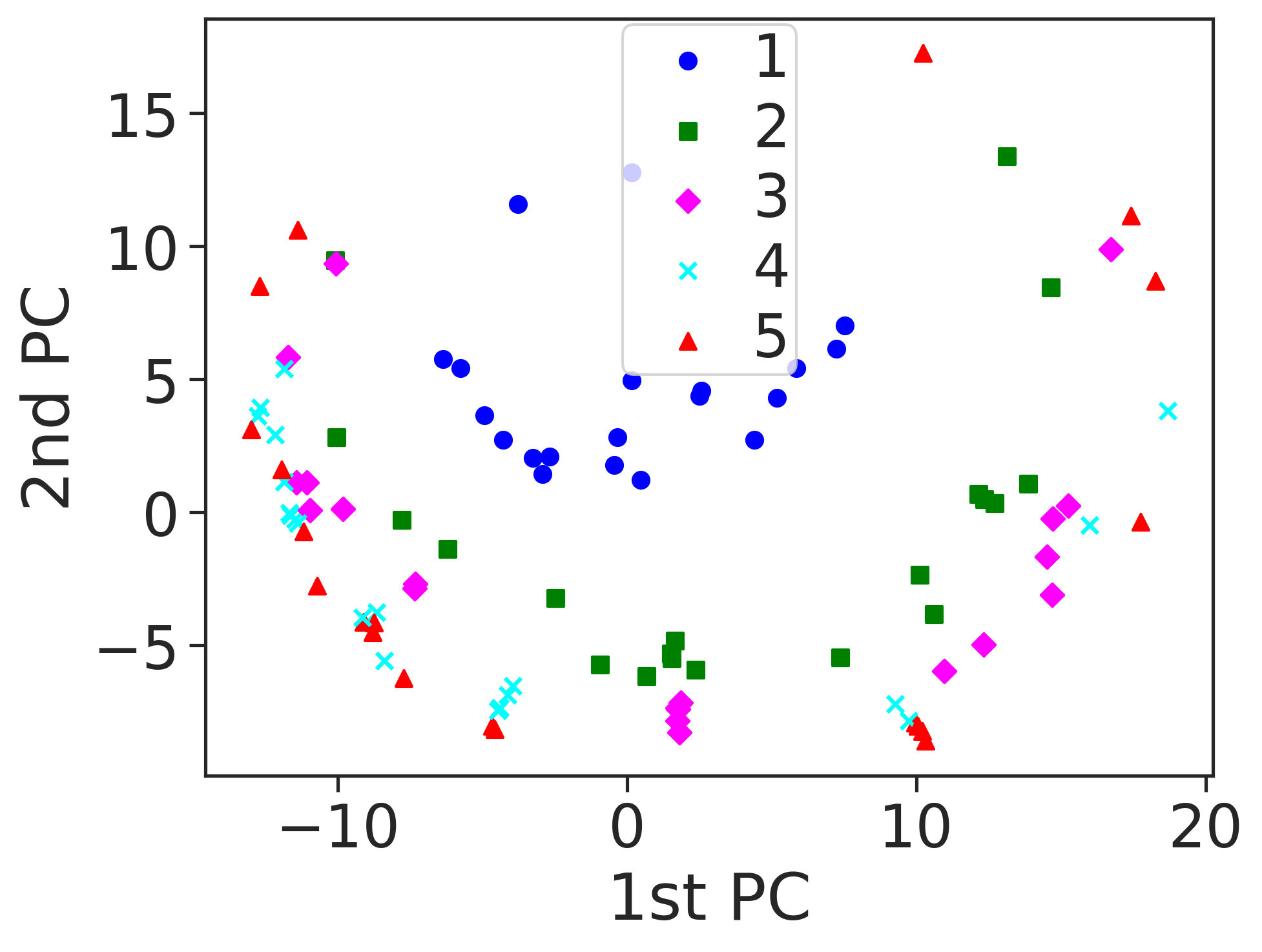}
    \hfill
    \includegraphics[width=0.3\textwidth]{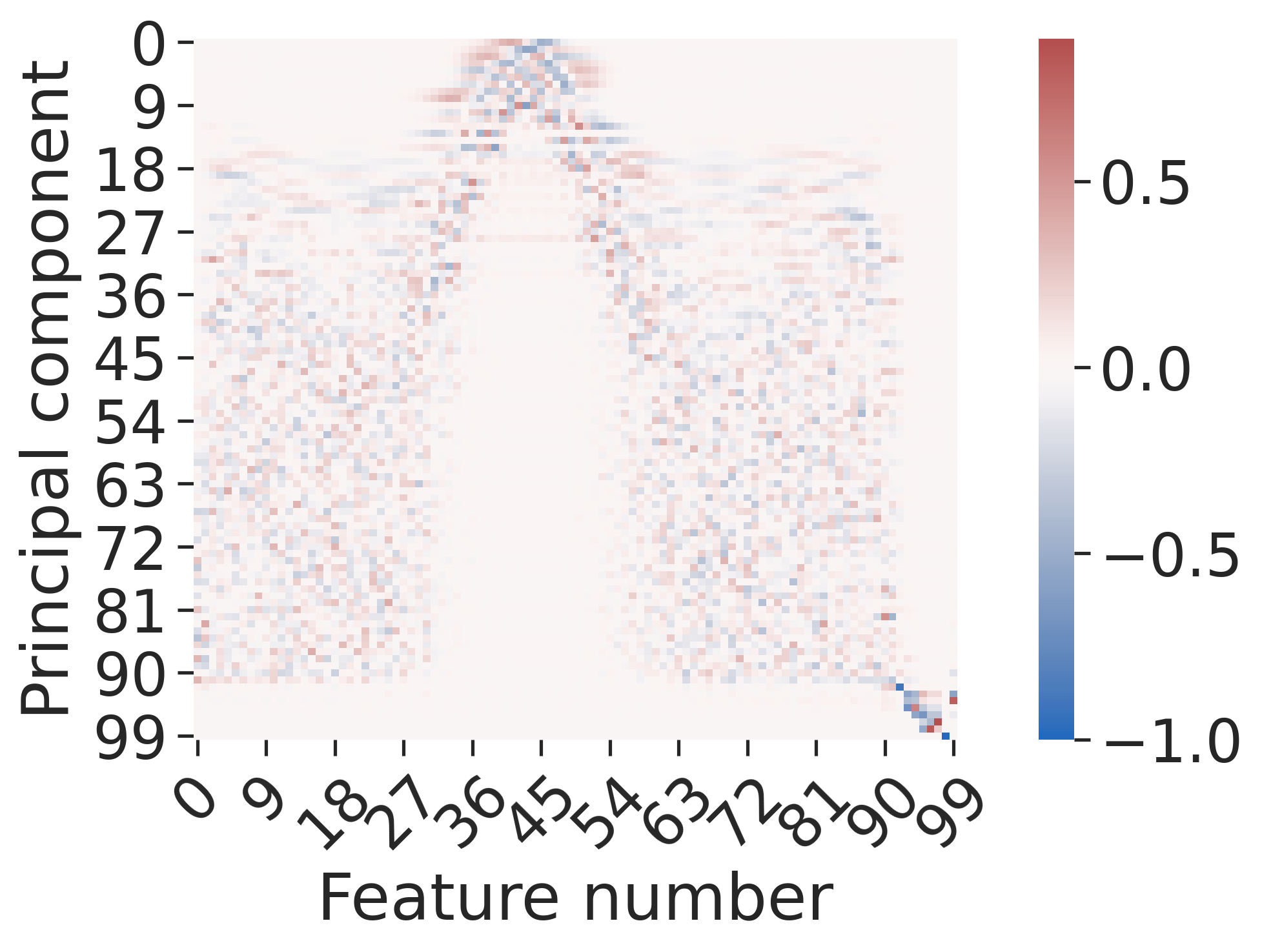}
    \caption{Wave packet signatures (left) for various realisations of a Erd{\H{o}}s–R{\'e}nyi ($N=50$, $\langle k \rangle=6$) graph with 5 and 1 layers (measured on one layer only). WPS projected onto a 2D space with PCA (centre) where colours and symbols distinguish the no. of layers. Each point is a different ER graph. PCA transformation matrix (right) showing the contributions of given WPS bins into principal components.}
    \label{fig:wps-er}
\end{figure*}

\begin{figure*}
    \includegraphics[width=0.3\textwidth]{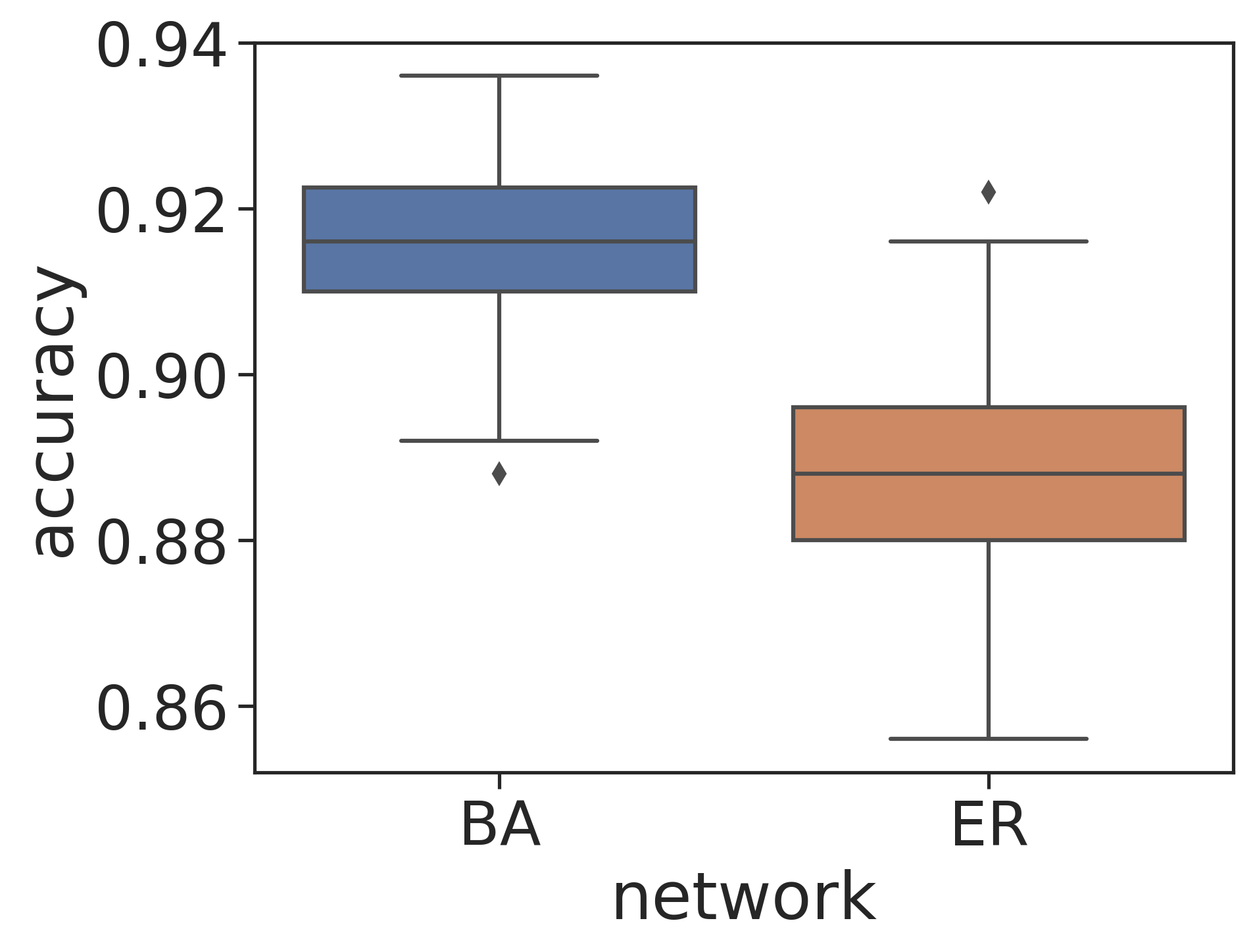}
    \hfill
    \includegraphics[width=0.3\textwidth]{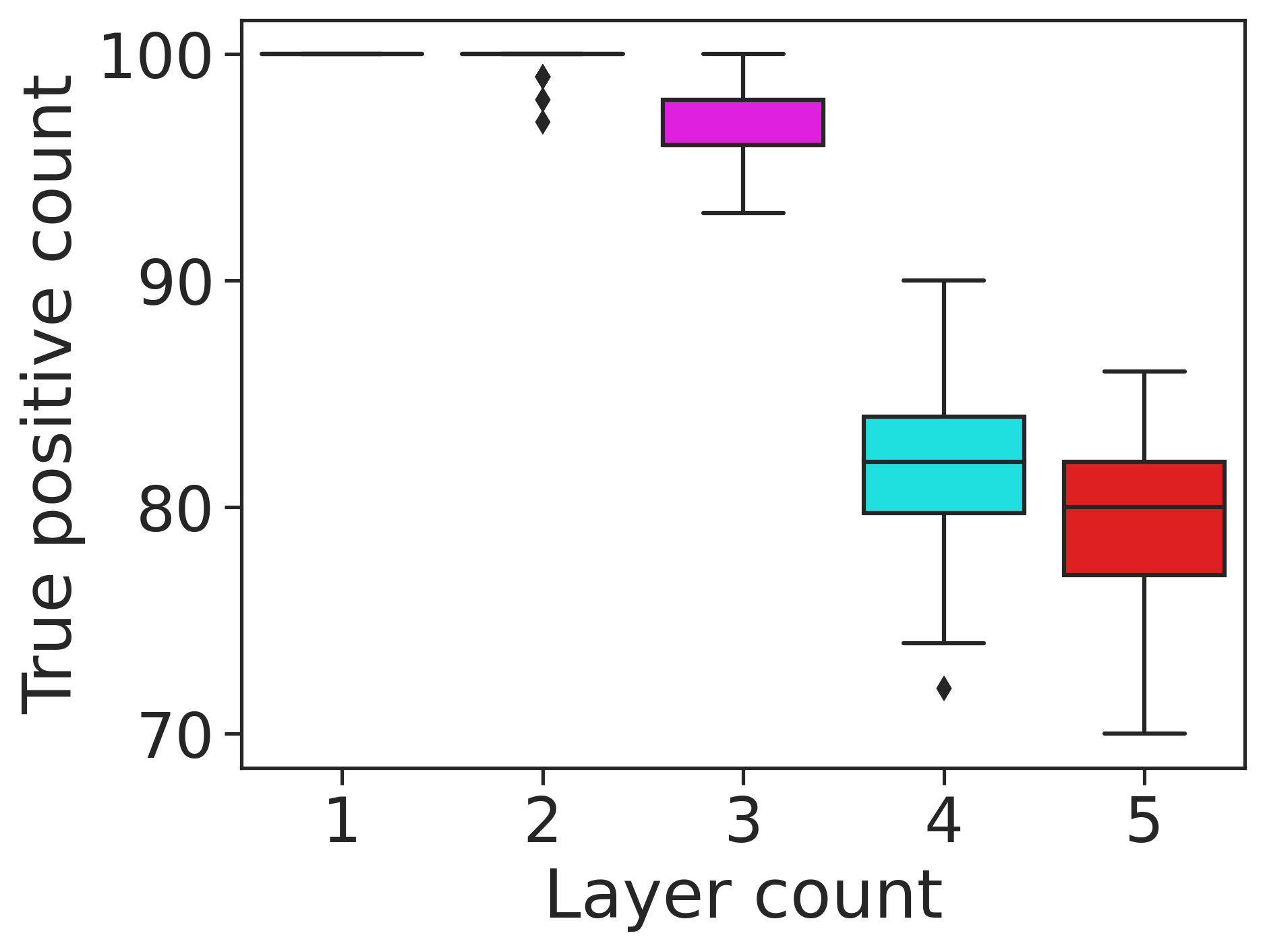}
    \hfill
    \includegraphics[width=0.3\textwidth]{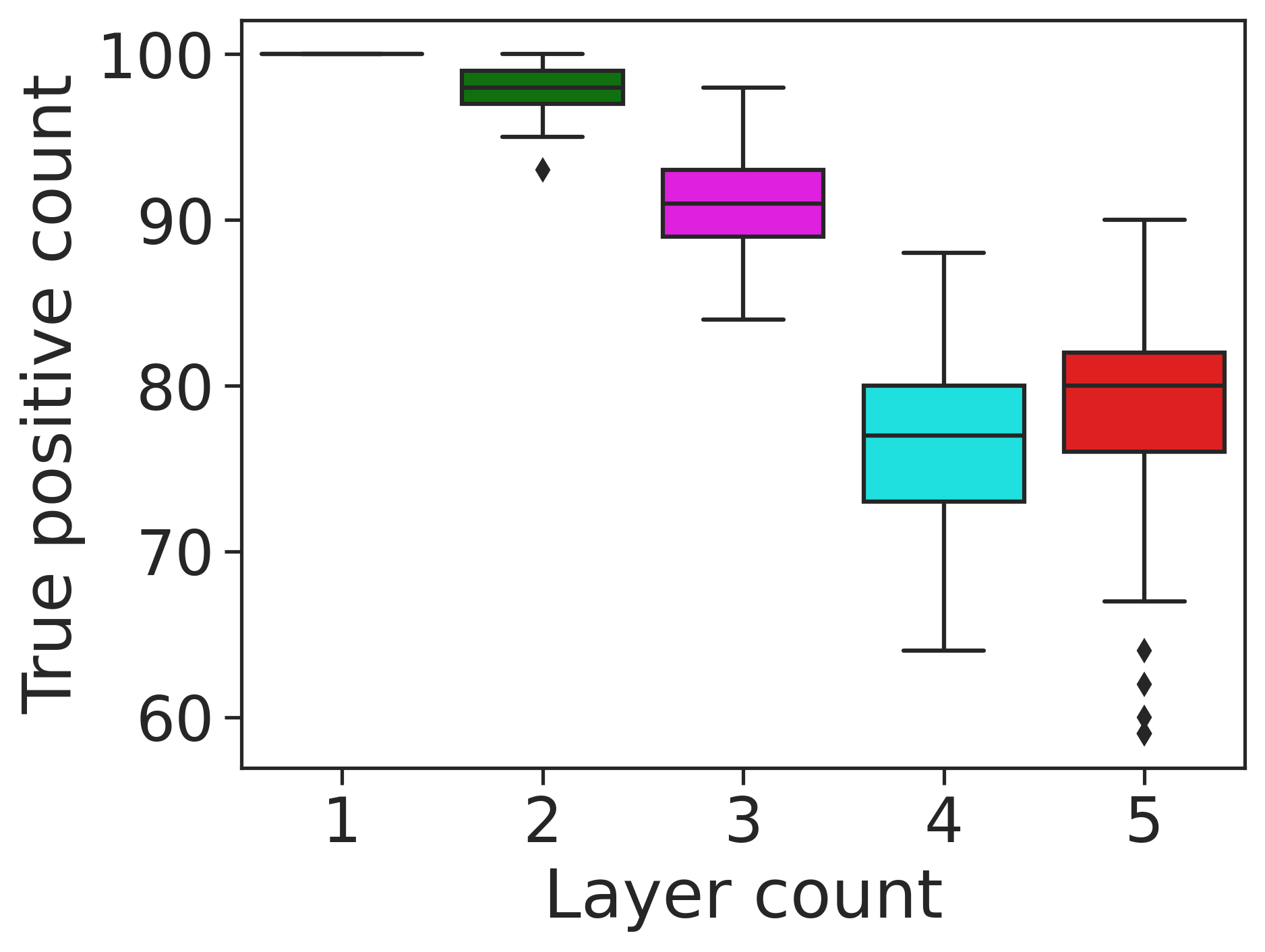}
    \caption{K-nearest neighbours classification accuracy of the layer count for a BA and ER graph as box plots (left) based on single layer WPSs. Contingency table diagonal values as box plots for a BA (centre) and ER (right). We simulated 400 independent realisations of a given graph type (BA, $N=50$, $m=3$; ER, $N=50$, $\langle k \rangle = 6$). For each type of graph we conducted 100 rounds of a 10-fold cross-validation to determine the K parameter in K-NN withdrawing 100 realisations (25\%) for the purposes of the final evaluation. The train/test split was random and stratified.}
    \label{fig:k-nn}
\end{figure*}

The rest of the procedure is similar, i.e., we create a WPS and we train a classifier on a given type of a graph (e.g., BA) and this time the classes are the number of layers. See Fig.~\ref{fig:wps-ba} and \ref{fig:wps-er} where we compare WPSs of a 5-layer BA graph vs 1-layer, and a 5-layer ER graph vs 1-layer, respectively. One can clearly see that the signatures are distinct. To further illustrate this we use the Principal Component Analysis (PCA, see Appendix \ref{app:b} for details) to project the WPSs onto a 2D space (see the centre pieces of Fig.~\ref{fig:wps-ba} and \ref{fig:wps-er}). There, one can see that each class takes a distinct region of space and thus we should be able to discriminate between them. However, it is worth noting that the more layers there are the more tightly packed the observations become, i.e., discriminating between a mono-layer and a penta-layer graph is fairly easy but between a tetra- and a penta- not as much. It is also worth noting that through the PCA we can see that the centre bins carry the most variance in the feature space (see panels on the right in Fig.~\ref{fig:wps-ba},~\ref{fig:wps-er}) and that there are certain distinct structures visible in the higher PCs for both BA and ER graphs.

As mentioned before, we follow Aziz \textit{et al.} and also choose the K-NN for classification (see Appendix \ref{app:c} for a description of the K-NN method). We build the model on various graphs with between 1 and 5 layers (with values measured only on one of the layers each time as explained earlier) and then test it to see if it can recognise how many layers there are in an unknown graph. We conducted our tests for the BA and ER graphs (see Fig.~\ref{fig:k-nn} for the results of classification). For each type of graph we simulated 500 independent realisations (100 per each number of layers) with the mean degree $\langle k \rangle = 6$ and network size $N=50$. We take 100 randomly chosen, in a stratified manner (i.e., classes are equally represented), realisations out of the data set - this will be the test set. On the training set we conduct 100 rounds of training and then testing. That is, in each round we do the training/test split, then a 10-fold cross-validation on the training set to find the best K parameter of the K-NN model and then use that on the test set. We present the results in a form of box plots. In Fig.~\ref{fig:k-nn} we show the accuracy of the classifier for both the BA and ER. We can see that while for the ER the results are slightly lower than for the BA, in both cases the accuracy is still fairly high with medians of 92 and 89 for BA and ER, respectively. 

To additionally illustrate the point made earlier that higher numbers of layers are more difficult to be discerned amongst one another, we show the diagonal values of the contingency tables from all 100 rounds for the BA (centre panel of Fig.~\ref{fig:k-nn}) and the ER (the right panel of Fig.~\ref{fig:k-nn}). One can clearly see that identifying mono-layer systems is practically 100\% accurate and it is the more layered systems that cause trouble for the classifier.

While the classification results seem very promising, it is important to note the obvious major disadvantage of this approach. One must build a training set for it to work. With synthetic networks (such as BA or ER graphs) it is easy to generate as many as one wants therefore creating an extensive training set and the limitation is purely in the computing power. When dealing with real-world networks one often does not simply have the ability to obtain similar enough graphs to the one currently under observation but with added layers --- each real-world network is unique. In such a circumstance perhaps a combination of many different synthetic networks obtained by introducing noise to a real-world network, i.e., rewiring some of the links, creating perturbed versions of the original network could suffice and other, more advanced, classification methods than K-NN could be utilised. That, however, is beyond the scope of this study.

\section{Fourier transform of the amplitude signal}
Here we introduce a new approach to detecting layers on quantum graphs. 
This completely novel way does not share the same issues as the previous one as it does not rely on building a classifier. Therefore, there is no need to build a training set and deal with all other aspects of the machine learning approach (like the choice of the classifier, tuning hyper-parameters etc.). While obviously not without its own problems, we believe that it is a more powerful and explainable tool while using very similar measurements as the previous method. 
Similarly to the previous case we assume we can either produce or observe a wave propagation on the graph initiated by a Gaussian wave packet. For efficiency's sake in the simulations we used the edge with the highest betweenness centrality as before, and we also measure the amplitudes at integer times in the centres of all edges for $2|E|$ times. While in the WPS method we simply follow the advice of Aziz \textit{et al.} for the count of measurements, in the approach discussed here it is usually rather clear if enough data was collected by a straight forward visual inspection.

Our proposition is as follows, at each integer time compute the sum of all amplitudes in the visible part of the system and treat it as a time dependent signal $S(t)$. Transform the signal into a frequency domain via a fast Fourier transform (see Appendix \ref{app:d} and \cite{cooley1965algorithm}) - $\hat{S}(f)$ - and look at the spectrum of the signal - $|\hat{S}(f)|^2$.

\begin{figure*}[tb]
     \centering
     \includegraphics[width=\textwidth]{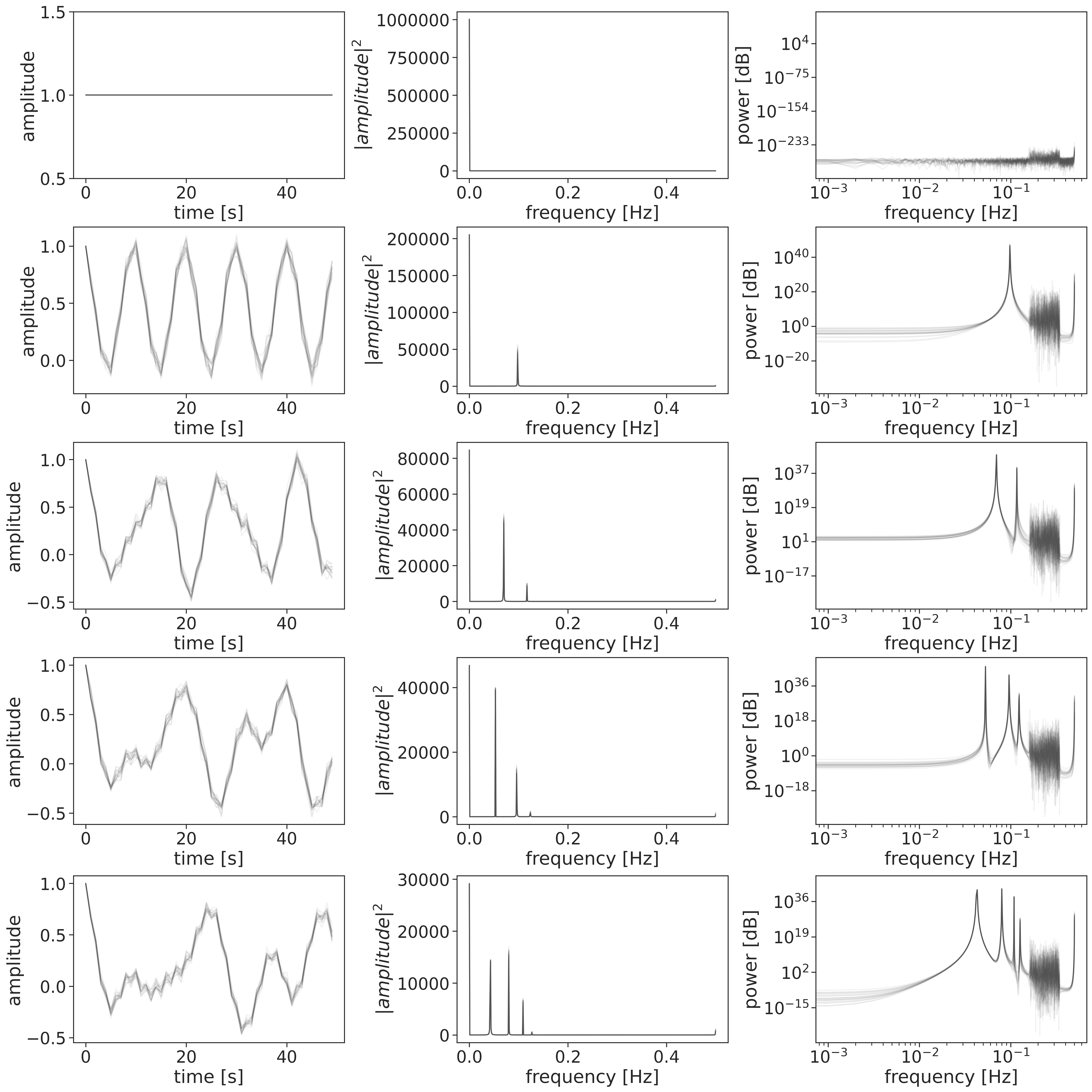}
     \caption{Sum of the amplitudes time evolution as measured on the only visible layer (left column). A fast Fourier transform of this signal (centre column) and its power spectrum (right column). Each row represents a different number of layers (1 to 5, going top to bottom). Simulations were conducted on 20 independent realisations of a BA graph ($N=50$, $m=3$) per row, overlaid with transparency. Note: for the purposes of the Fourier analysis we use signals of length $10^3$ s.}
     \label{fig:fft-ba}
 \end{figure*}

A mono-layer system will produce a ``flat'' signal $S(t)$, whilst a multi-layer one will exhibit periodic behaviour due to the energy leaking in and out of the visible layer from and into other layers. At a sufficiently long measurement time the signal should stabilise and become stationary as long as no perturbation is introduced to the overall network. This transfer of energy induces oscillations in the amplitude sum on the visible layer which in turn create clear peaks in the power spectrum, see Fig.~\ref{fig:fft-ba} for the results from the BA graphs (and Fig. \ref{fig:fft-er} for the ER, presented in Appendix~\ref{app:er}). The left column shows the signal $S(t)$, centre $|\hat{S}(f)|^2$, and right $|\hat{S}(f)|^2$ in decibels and a log-log scale. Each row has 20 independent realisations plotted on top of each other, with transparency, to show that these peaks are fairly consistent, and different number of layers (1st row are mono-layer systems, 2nd row bi-layer, etc.). It is quite apparent that there are visible peaks and their count strictly corresponds to the number of layers in the system.

These results already show the advantage of FFT over WPS as it is simpler and does not seem to suffer from struggling to differentiate between highly layered systems as much. Additionally, it does not require any prior knowledge or learning of the model. It is far from flawless, however. As it is much easier to test on real networks than WPS we applied it to three real-world networks \cite{vickers1981representing, krackhardt1987cognitive, chen2006wiring} - see descriptions in Appendix \ref{app:e}. The results are in Fig.~\ref{fig:fft-real} where one can see that while we do get peaks in the spectrum and therefore can confidently state that there are hidden layers, it is much less clear how many of them there are. This is most likely due to the fact that these networks are not as clear cut multiplexes as the synthetic systems we discussed earlier. These networks here have various mean degrees in each layer and that also implies varied coupling strength between the layers. Moreover, real-world networks can have other characteristics that differ between the layers such as clustering coefficients or degree distributions, and the synthetic systems tested simply do not have this property. However, we show in the next section that it is possible to reconstruct the full spectrum of the row-normalised adjacency matrix with this method and therefore determine the number of layers for any system. 

\begin{figure*}[tb]
     \includegraphics[width=\textwidth]{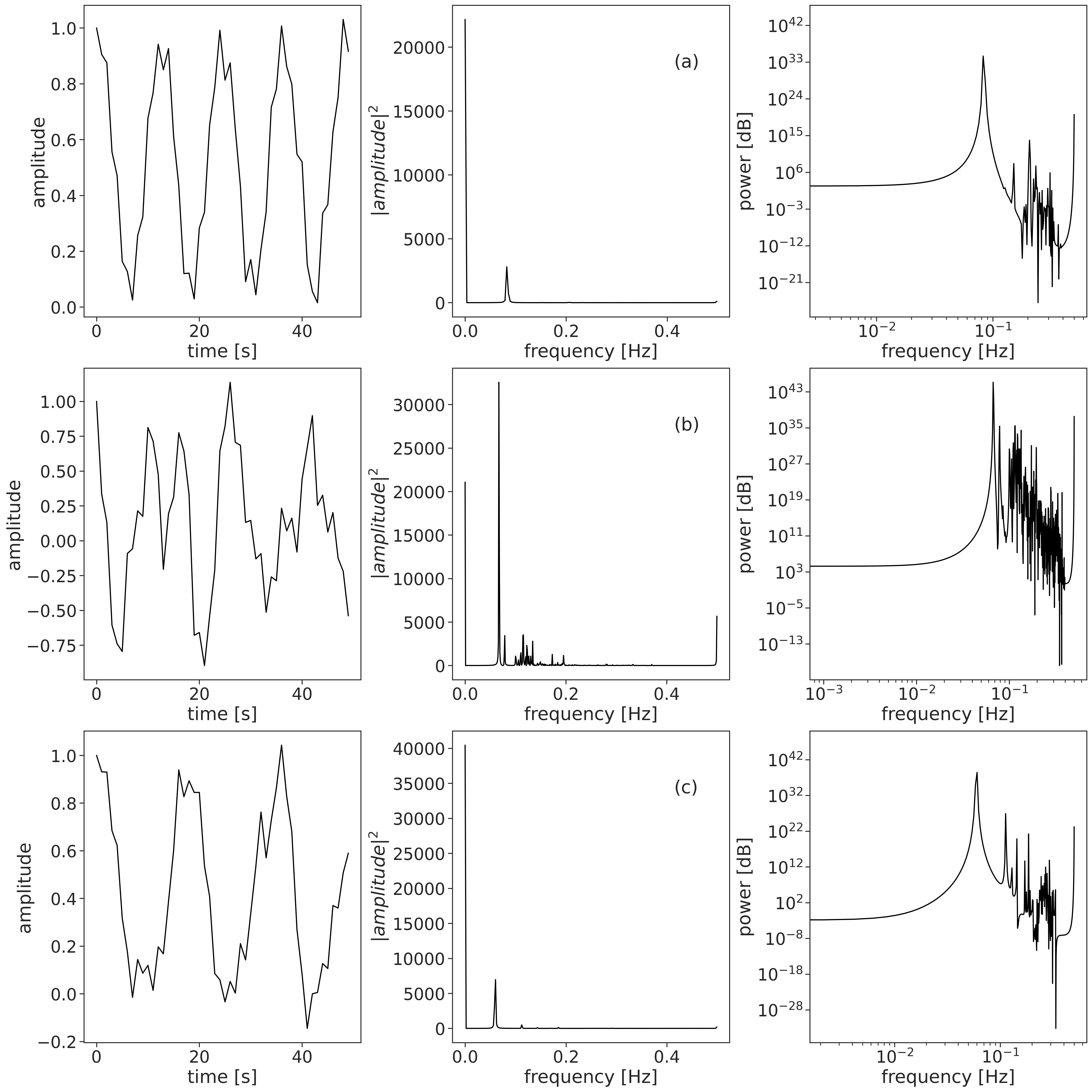}
     \caption{Sum of the amplitudes time evolution as measured on the only visible layer (left column). A fast Fourier transform of this signal (centre column) and its power spectrum (right column). Each row represents a different real-world network (as indicated by the labels -- (a) Vickers \cite{vickers1981representing}, (b) C. Elegans \cite{chen2006wiring}, (c) Krackhardt \cite{krackhardt1987cognitive}). Each graph is a 3-layered multiplex. Note: for the purposes of the Fourier analysis we use signals of length $10^3$ s.}
     \label{fig:fft-real}
 \end{figure*}

\section{Spectrum reconstruction}
In this section we show that with a sufficiently long observation it is possible to completely recover all eigenvalues of the row-normalised adjacency matrix of the full system and thus trivially determine the number of hidden layers.

We shift slightly from the previous sections as we no longer take the measurements on the edges but only on the nodes. This makes the problem less computationally intensive and also, in our opinion, makes for a more practical case as it might be sometimes easier to observe just the nodes' states (see Appendix \ref{app:a} for details). However, the same analysis here can be applied using edge measurements and the one in the previous sections could be done with the node values only - we chose otherwise as we are stemming from the work of Aziz \textit{et al.}

Similarly as before we observe the sum of amplitudes, however, in this case it is important to have enough samples of the signal to provide sufficient resolution in the power spectrum. How long one needs to observe a system will of course depend on the intricacies (and mostly size) of the system in question. 
As the propagation process is not stochastic the time needed for observation is finite and in our experience not unattainable. 
The goal is simply to have the complete power spectrum of the signal. Then one takes note of the peaks present - we opted for an automated approach using a wavelet transform \cite{peakdet}(see Appendix \ref{app:f}). 
These peaks in the power spectrum correspond to the eigenvalues of the Hamiltonian (divided by $2\pi$) which in turn directly relate to the eigenvalues of the row-normalised adjacency matrix - $\hat{\mathbf{A}}$ - such that each eigenvalue $\lambda \notin \{-1, 1\}$ of $\hat{\mathbf{A}}$ has the corresponding Hamiltonian eigenvalues $\arccos(\lambda)$ and $2\pi-\arccos(\lambda)$.
This leads us to two important results, (i) the number of frequencies present in the power spectrum $\#f$ is two less than the count of eigenvalues of the adjacency matrix and since we know the layer sizes (i.e., node count per layer - $N$) as a multiplex structure was assumed, the number of layers $K = (\#f + 2) / N$; (ii) we can in fact recover almost exactly all eigenvalues of $\hat{\mathbf{A}}$ as $\cos(2\pi f_i)$ for each frequency peak $f_i$ in the power spectrum.

\begin{figure*}[tb]
     \includegraphics[width=0.3\textwidth]{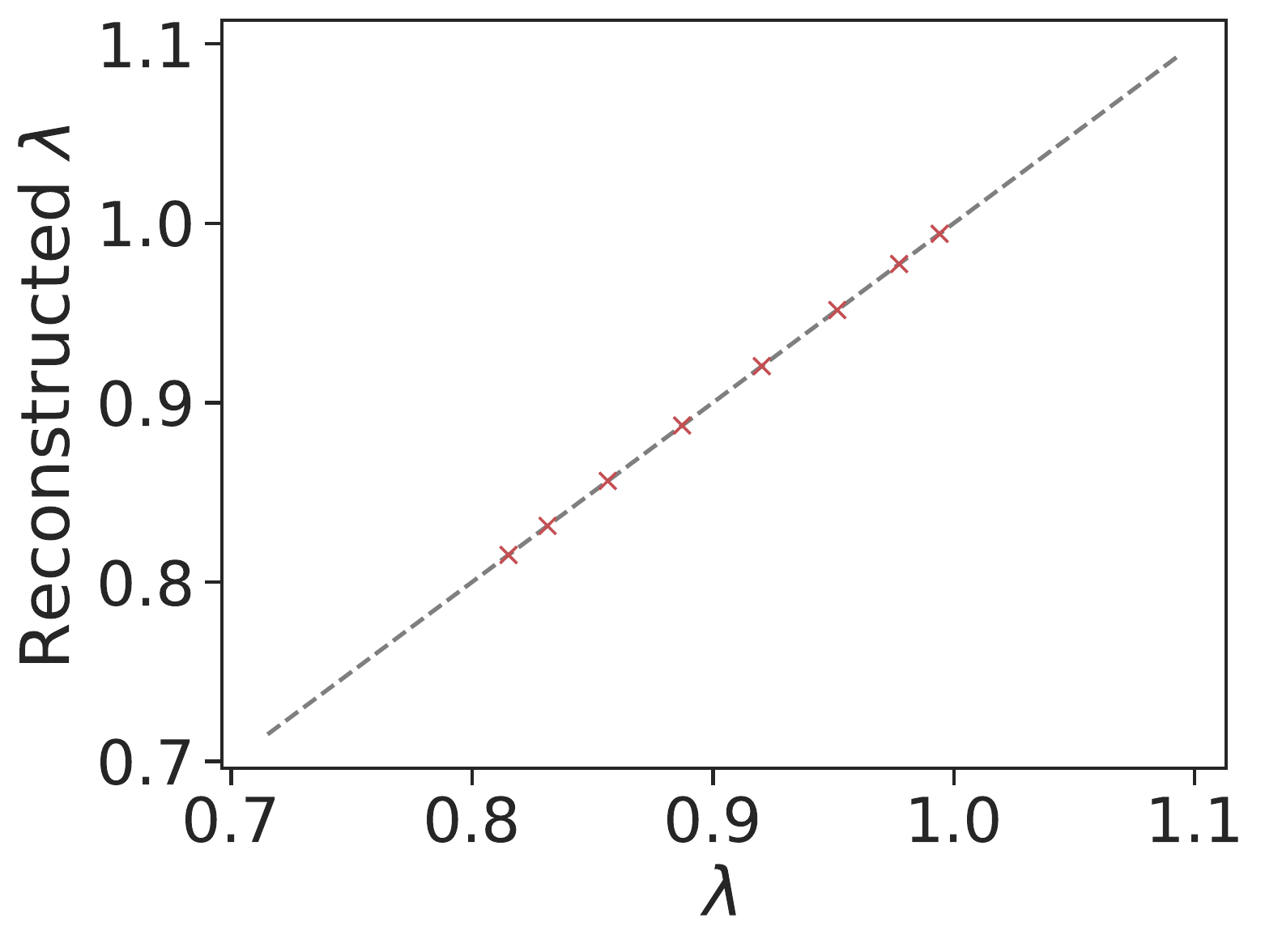}
     \hfill
     \includegraphics[width=0.3\textwidth]{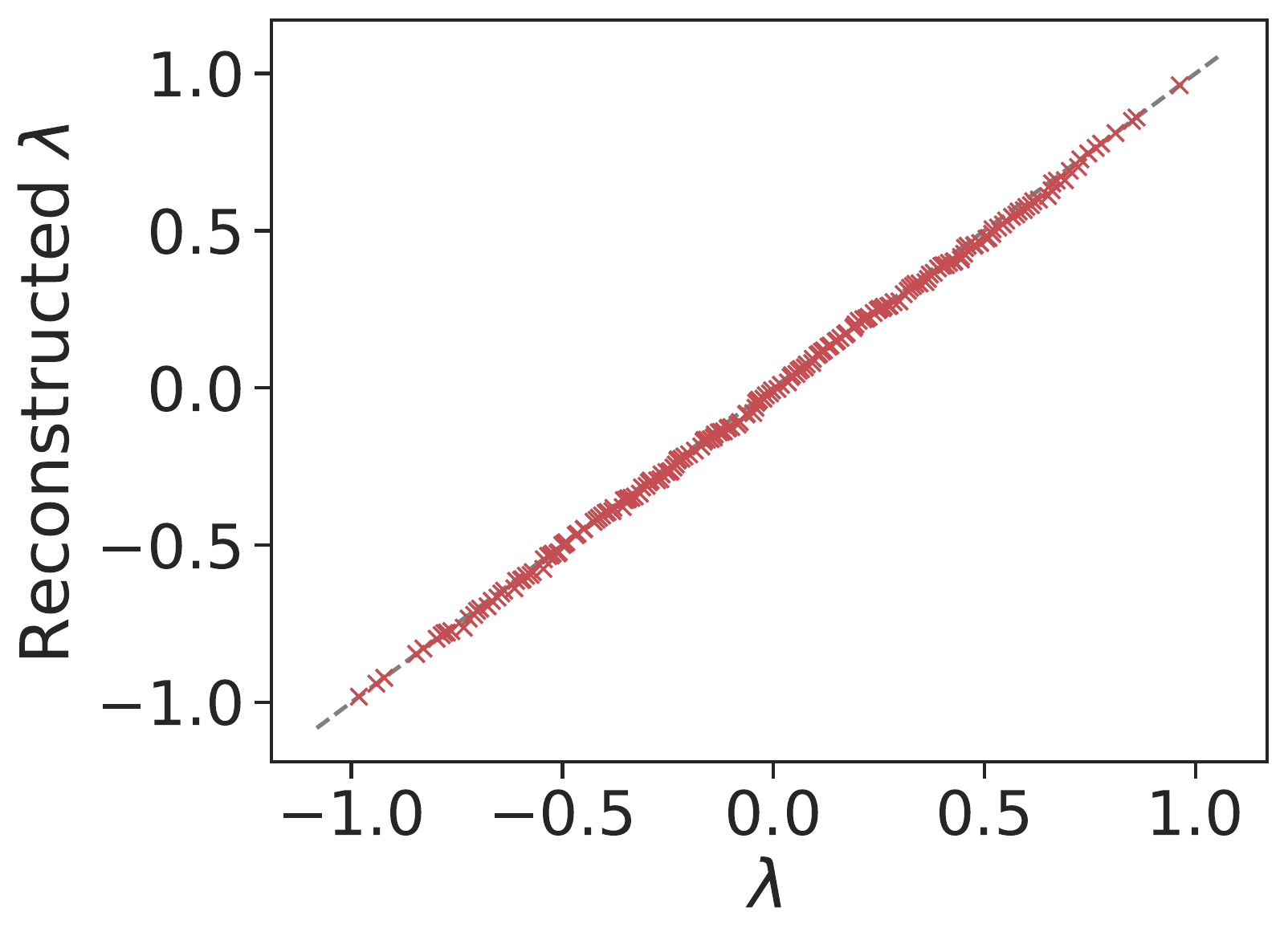}
     \hfill
     \includegraphics[width=0.3\textwidth]{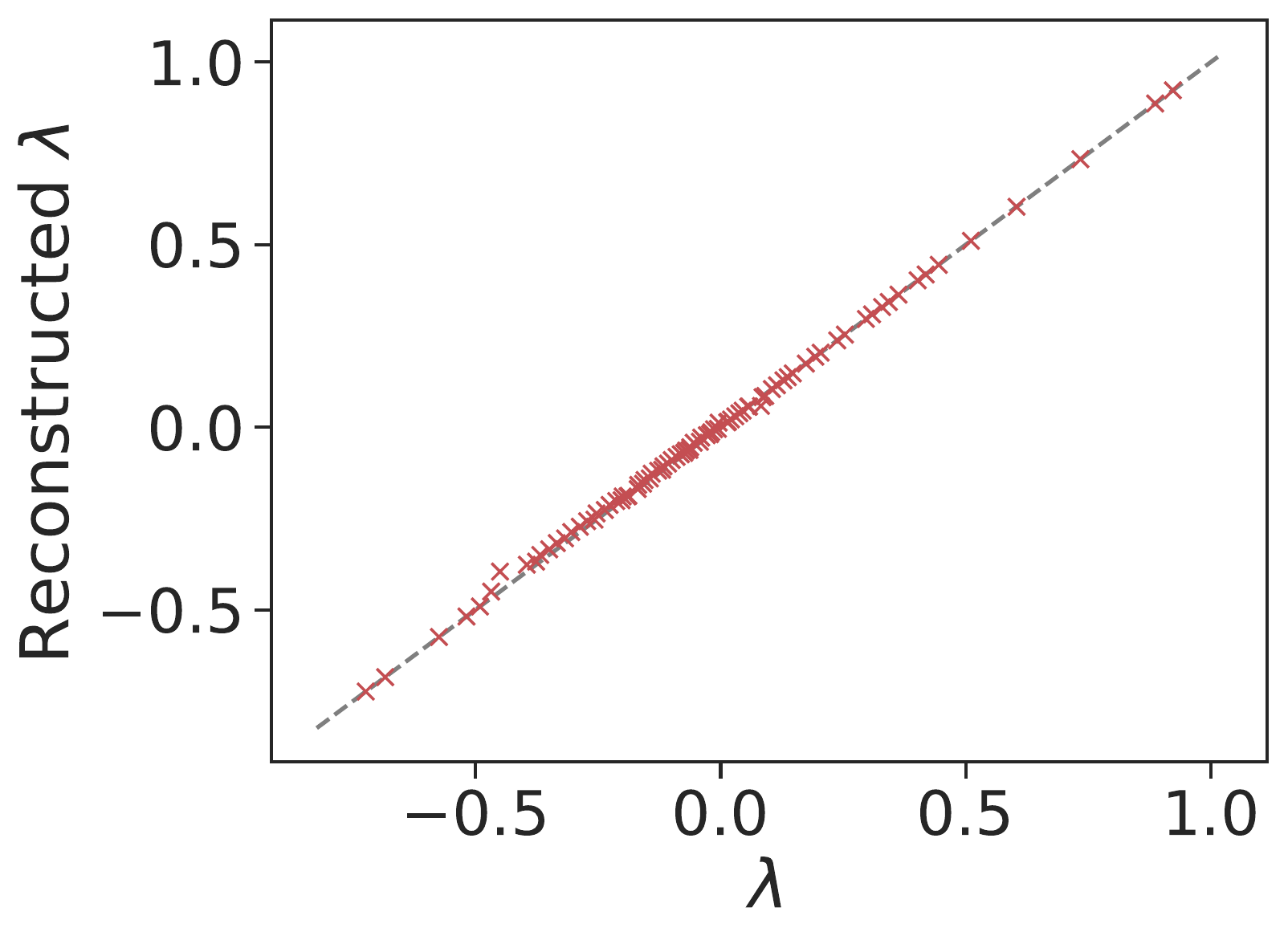}
     \caption{Eigenvalues reconstruction from the Fourier spectrum of the nodes' amplitudes sum signal. (Left) an example of a complete graph multiplex with a layer size $N=20$ and number of layers $K=9$. As it is a special case of an extremely symmetric adjacency matrix there are only as many eigenvalues as there are layers. (Centre) a Barab{\'a}si-Albert graph with $N=50,~m=3,~K=4$ has a much more complex spectrum and so does a real-world network (right) - Vickers\cite{vickers1981representing} - for both of which we attain an almost perfect match between the recovered and actual eigenvalues. The dashed diagonal line is a visual aid showing ``$y=x$''.}
     \label{fig:reconstruction}
 \end{figure*}

We present the result of the full spectrum reconstruction in Fig.~\ref{fig:reconstruction} for a complete, BA and real-world graph. We chose the complete graph as it has a special case due to the extreme symmetries of the multiplex adjacency matrix and thus the number of peaks directly corresponds to the number of layers unlike more complex cases where the eigenvalues' multiplicities behave differently. Of course, this does imply that if due to some particular structures in a given system some eigenvalues have high multiplicity, the simple formula $K = (\#f + 2) / N$ will not hold and system specific adjustments would be needed.
The reconstructed eigenvalues give an almost perfect match with those of the row-normalised adjacency matrix. Note that the performance here is mostly limited by the peak detection method and the resolution in the frequency domain, i.e., the information is there in the spectrum, the only challenge is to recover it efficiently. 

\begin{figure*}[tb]
     \centering
     \includegraphics[width=.75\textwidth]{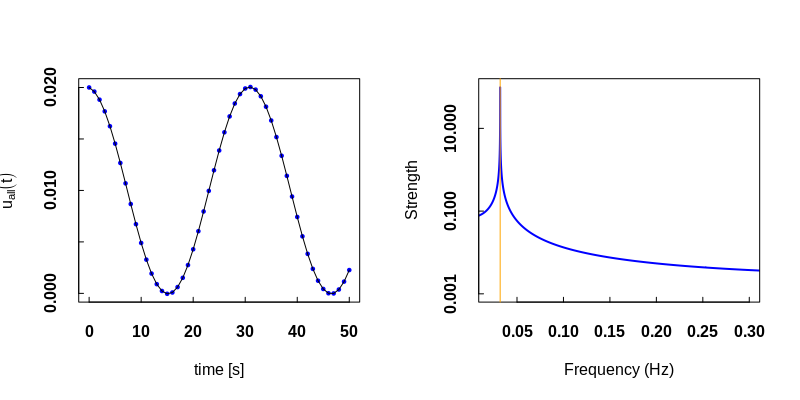}
     \caption{Full graph duplex with $N=100$ nodes in each layer: (left) the sum of amplitudes over time in a single layer; lines are guidance to eye connecting theoretical predictions from Eq. (\ref{eq:ualldup}) that are exactly covered with points obtained from numerical simulations, (right) the power spectrum of the signal shown in the left panel [the series consists of 10000 elements]. The vertical line is the frequency equal to $\omega_{2N-1} / 2\pi = (N-2)/(2\pi N)$.}
     \label{fig:ualldup}
 \end{figure*}

In order to explicitly show the correspondence between the eigenvalues of the Hamiltonian and the peaks of the power spectrum we follow in detail the full graph case, considering the simplest configurations -- a monoplex and a duplex network (see Appendix \ref{app:full}). Our exact analytical solutions prove that for the duplex network of $N$ nodes on each layer, among $2N$ eigenvalues that characterise the system we have only four distinct ones $\bm{\lambda} =\left\{-\frac{2}{N},0,\frac{N-2}{N},1\right\}$, the first two having multiplicity of $N-1$. Figure \ref{fig:ualldup} shows that in such a system one recovers just one eigenvalue, i.e., $\lambda = \frac{N-2}{N}$ directly connected to periodicity of the sum of observed amplitudes on the nodes. On the other hand, observing a single node allows for the recovery of the full spectrum (see Appendix \ref{app:full} and Fig. \ref{fig:undup} for details). 

\section{Discussion}
In this paper we explore the paradigm of quantum graphs as a potential tool for studying multi-layer networks.
The particular problem we are interested in is determining whether there are hidden layers of communication in the system, that we cannot fully observe, by taking measurements of the ongoing dynamics in a single layer that we can observe.
We proposed and tested two methods - one based upon a Gaussian wave packet signature (WPS) that was introduced previously by Aziz \textit{et al.} to discriminate between various types of mono-layer systems, and the other on observing the power spectrum of the wave amplitudes.

WPS is a method where a Gaussian wave packet, either observed or purposefully produced, initiates the propagation from a single edge and we take the measurements of the amplitudes at every edge at integer times for a sufficiently large number of times. Such data is then histogrammed to produce the signature. This signature has the property of being similar within a category of graphs while varied without. That is, e.g., signatures of the ER graphs are similar to one another but different from the BA graphs, or for our purposes, mono-layer graphs have different signatures than multi-layer ones etc.
This in turn can be utilised by machine learning models, such as K-nearest neighbours, to build a model capable of discriminating between graphs with different numbers of layers. This approach suffers from several issues, however. Most prominently, it requires a training sample. This can be very difficult to obtain in real-world scenarios and while perhaps a well varied synthetic data set could suffice, at this point it is mere speculation. The choice of an appropriate machine learning scheme and its construction is also a non-trivial task. Additionally, as the number of layers grows, differentiating between such networks becomes increasingly difficult since the signatures become less varied.

We also introduce an approach that utilises a discrete Fourier transform (DFT) instead of a machine learning model. Instead of histogramming the measurements as before, one computes the sum of the amplitudes on the visible layer at each integer time. This constitutes a signal that at sufficient time scales should become stationary. In a mono-layer system the signal will simply be a constant value due to energy conservation. However, should other layers be present in the system, from the perspective of the mono-layer there will be oscillations as the energy will flow out and back into it. We can inspect those oscillations with the use of the DFT and look at the power spectrum. The spectrum will exhibit characteristic peaks absent in the mono-layer networks. The number of these peaks strictly corresponds with the number of layers in the synthetic scenarios tested. This approach is significantly advantageous over the WPS as it does not require building a learning sample and is in general much simpler. Furthermore, it does not really suffer in terms of differentiating, e.g., tetra- from penta-layer systems, etc. Although it shows much promise in synthetic scenarios, it does not perform as well in real-world networks. It does indeed indicate clearly that there are hidden layers but the number of them can be rather tricky to discern. This is perhaps not that surprising considering that real-world networks are much more ``messy'' in some way than synthetic examples. Layers vary in size, degree distribution, clustering coefficients and so on and so forth, while, e.g., a penta-layer BA graph shares all characteristics between layers even though the exact connections are different. Those and other features of real-world systems could also affect the coupling amongst the layers that most certainly will affect the nature of the amplitude signal.

In such cases (i.e., where peak count after a brief observation is not enough) we show that simply a longer observation time is required. As the signal is not stochastic and oscillation periods are finite it does not seem unfeasible to observe enough of the signal to determine its power spectrum with sufficient resolution. Then each peak in the spectrum corresponds to the Hamiltonian eigenvalues that in turn are related to the eigenvalues of the row-normalised adjacency matrix via a simple formula. With this we showed that it is indeed possible to recover all these eigenvalues and thus trivially determine the number of layers in the system.

It is worth underlining here that a row-normalised adjacency matrix is in fact the so called right stochastic matrix of a given graph and while it goes beyond the scope of this paper, there exist methods of reconstructing the whole matrix from its spectrum \cite{chu2005inverse, steidl2020new, cacace2019karpelevich, zhao2016geometric, orsi2006numerical} which we suspect should be quite feasible considering we already assume knowing part of it (one layer and inter-layer structure). That in turn could also open the door to the adjacency matrix itself. Recovering all the connections exactly may not be possible, however, having a matrix iso-spectral to the adjacency matrix is also very valuable as having this spectrum allows for determining many important properties of the system \cite{sanchez2014dimensionality, cozzo2016multilayer}.

The issue of -- how long the observation time should be -- still remains, however. In the experiments presented in this paper we needed more observation points using the Fourier approach than with the machine learning one.
In the real-world example used in this article the necessary measurement time to get the almost perfect match seen at the right panel of Fig.~\ref{fig:reconstruction} is fairly substantial  -- we simulated the system for $11520$ time steps to be exact. While lower values would not mean no match at all, the low power components do become less discernible. This is due to the fact that some frequencies may be difficult to observe because of the particular intricacies of a given system, while some eigenvalues are going to be detectable more easily since more power is associated with them.
In order to observe all frequencies, in an idealised scenario, we should observe at least two periods of the signal (one could also possible suffice but we shall assume two). Therefore, the total observation time should be $4 \pi / \arccos(\lambda_2)$, where $\lambda_2$ is the eigenvalue with the second largest magnitude of the stochastic matrix.
In general, i.e., for any system, for any stochastic matrix, it is not possible to give one formula for the behaviour of $\lambda_2$\cite{bialas1998hypothesis}. In special cases it is possible to provide an analytical expression and, e.g., for a full graph duplex we would need $4 \pi / \arccos((N-2)/N)$ measurements which is clearly much fewer than in case of the machine learning approach -- $N(N-1)$ (note that it does not hold for networks like Vickers). The latter has also been rather arbitrarily chosen by Aziz et al., and further more, one could also ask how big the training test must be to achieve a certain level of accuracy. This, similarly to assessing the necessary number of measurements, is rarely, if ever, possible to be known a priori. 
This points to an another possible advantage of our Fourier approach (despite requiring more observation time in some systems), which is that since we do not require training of the model we can in fact start recovering the spectrum ``online'', i.e., with very few measurements and simply keep adding them and improving the results as the time passes. As we continue our measurements it will also usually become clear which frequency peaks are stable and whether all of them are already expressed via a visual inspection of the signal or the spectrum.

We find that both methods presented in this paper - the WPS and FFT - show enough success in these simple scenarios we tested to merit further study, such as in noisy (or stochastic) systems, for instance. They each have their pros and cons that we hopefully managed to outline clearly as well as the potential room for improvement of their applicability and understanding of waves on quantum graphs alike.

While we find our approach to be very promising it is also important to outline the potential further study areas as we could not possibly have covered everything in this article. It would be of substantial interest to investigate how the described system behaves when distinct community structures (as in, e.g., a stochastic block model \cite{HOLLAND1983109}) are present. We suspect that it is feasible to detect the communities themselves, however, discerning these communities from layers can pose an additional challenge. Furthermore, we have tested a particular form of multi-layer networks with distinct symmetries, yet other, more complicated types exist, such as networks of networks \cite{aleta} and multi-layer graphs with complex inter-layer connections \cite{boccaletti2014structure}. These are especially interesting and considerably more difficult as the inter-layer structures themselves can vary wildly. Finally, there is also a matter of node membership in specific layers. Depending on the particulars of the studied real-world scenarios it can be of interest to be able to ascertain where a given node belongs. We suspect that this may also be possible, as from our experience single-node time series can also contain a substantial amount of information. We hope to address these issues in our future work.

\begin{acknowledgements}
The project was partially funded (\L{}G) by POB Research Centre Cybersecurity and Data Science of Warsaw University of Technology within the Excellence Initiative Program - Research University (ID-UB). JS acknowledges support by National Science Centre, Poland Grant No. 2015/19/B/ST6/02612 while JAH was partially supported by the Russian Science Foundation, Agreement No 17-71-30029 with co-financing of Bank Saint Petersburg, Russia.
\end{acknowledgements}

\appendix
\section{Calculation of the wave amplitude}\label{app:a}
\subsection{Overall description}
The general solution of the wave equation on a graph $G$, provided that the initial condition is a Gaussian packet fully localised on a given edge $f$ is derived and presented in detail by Aziz \textit{et al.} in \cite{aziz_wave_2019}. Here, we re-write it in terms of integer times, i.e., $t=0,1,2,...$ so that it fits the case examined in the main text. We additionally assume that the graph is unweighted, undirected and non-bipartite. In such a setting, we consider an arbitrary edge $e=\{u,v\}$ that connects two vertices $u$ and $v$, and can be associated with a variable $x_e \in [0,1]$ that represents the coordinate along such an edge. Then the amplitude $u$ of the wave in the middle of the edge $e$ can be expressed as  
\begin{equation}
    u(e, f, t) = u_1(e, f, t) + u_5(e, f, t) + \frac{1}{|E|},
\end{equation}
where $|E|$ is the number of edges in the graph and $u_1$ and $u_5$ are defined as follows
\begin{widetext}
\begin{equation}
\begin{array}{l}
     u_1(e,f,t) = \sum\limits_{\omega \in \Omega} C(e,\omega) C(f,\omega) \cos(B(e,\omega)+\frac{1}{2}\omega)\cos(B(f,\omega)+\omega(\frac{1}{2} + t))\\
     u_5(e,f,t) = 2\cos(\pi t)\sum\limits_{i} C_{\pi}(e,i) C_{\pi}(f,i) 
\end{array}
\label{eq:u}
\end{equation}
\end{widetext}

In the above equations $\omega$, $C(e,\omega)$ and $B(e,\omega)$ come from the edge-based eigenvalues and eigenfunctions, which are, respectively $\omega^2$ and $\phi(e,x_e)=\pm C(e,\omega)\cos(B(e,\omega)+\omega x_e)$ of the row-normalised adjacency matrix $\hat{\mathbf{A}}$ of the graph $G$. Assuming that we know the vertex-based eigenvector-eigenvalues pairs ($g(v)$,$\lambda$) of the matrix $\hat{\mathbf{A}}$ we can express $C(e,\omega)$ and $B(e,\omega)$ as
\begin{equation}
\begin{array}{l}
     C(e,\omega)^2 = \frac{g(v,\omega)^2 + g(u,\omega)^2 - 2g(u,\omega)g(v,\omega)\cos\omega}{\sin^2\omega}\\
     \tan B(e,\omega) = \frac{g(v,\omega)\cos\omega - g(u,\omega)}{g(v,\omega)\sin\omega}, 
\end{array}
\label{eq:cb}
\end{equation}
while $\omega = \arccos{\lambda}$. The sign of $C(e,\omega)$ needs to chosen in order to match the phase, in practice it can be achieved by calculating $\mathrm{sgn}[g(v)]|C(e,\omega)|$. It is always true that one of the eigenvalues is equal to 1 (consequently, $\omega = 0$): this value is responsible for the constant term $1/E$ in Eq. (\ref{eq:u}) so it is not included in further calculations, i.e., it does not belong to the $\Omega$ set in the function $u_1$. Although $\phi(e,x_e)$ are orthogonal, they still need to be normalised. To fulfil this condition for each $\omega \in \Omega$ we calculate the normalisation factor $\rho(\omega)$

\begin{widetext}
\begin{equation}
    \rho(\omega) = \sqrt{\sum_{e}C(e,\omega)^2\left[ \frac{1}{2} + \frac{\sin(2\omega + 2B(e,\omega))-\sin(2B(e,\omega))}{4\omega}\right]}
\label{eq:rho}
\end{equation}
\end{widetext}
where $e$ runs over all edges in the graph $G$. Then, in order to obtain a properly normalised value of $C(e,\omega)$ one needs to divide it by $\rho(\omega)$.

For calculations of $C_{\pi}$ one first needs to transform the original undirected graph $G$ into a directed one $D(G)$ by simply replacing each edge $e=\{u,v\}$ with two arcs $(u,v)$ and $(v,u)$. In the next step we create a structure called the oriented line graph ($OLG$), constructed by substituting each arc of $D(G)$ by a vertex (such vertices are connected if the head of one arc meets the tail of another arc). Using the adjacency matrix $\mathbf{A}_{olg}$ of the $OLG$ we solve the eigenproblem $\mathbf{A}_{olg}\mathbf{g}_{olg}=\lambda_{olg}\mathbf{g}_{olg}$ and then restrict ourselves to $\lambda_{olg}=-1$ and the corresponding eigenvectors (there should be exactly $|E|-|V|$ linearly independent solutions) that form $C_{\pi}$. 

Let us note that if we decide to measure the wave amplitude on the nodes instead of the edges the formula is particularly simple as then
\begin{equation}
    u_n(e, f, t) = u_1(e, f, t) + \frac{1}{|E|},
\end{equation}
and
\begin{widetext}
\begin{equation}
     u_1(e,f,t) = \sum\limits_{\omega \in \Omega} C(e,\omega) C(f,\omega) 
     \left[\cos\left[\left(t - \frac{1}{2}\right)\omega - B(e,\omega) + B(f,\omega)\right] + \cos\left[\left(t + \frac{3}{2}\right)\omega + B(e,\omega) + B(f,\omega)\right]\right]
\label{eq:un}
\end{equation}
\end{widetext}
If follows that in such a case we do not have to create $OLG$ and thus the calculations are both less time and resources consuming.

\subsection{An example}
\begin{figure*}[tb]
     \includegraphics[width=.75\textwidth]{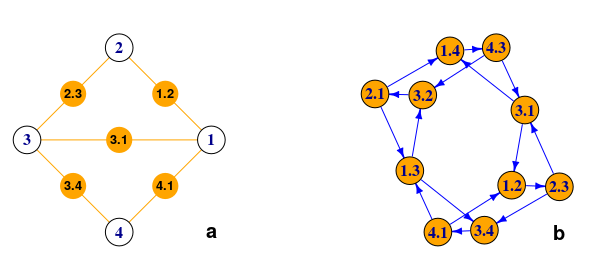}
     \caption{(a) An example of a simple graph consisting of $|V|=4$ nodes and $|E|=5$ edges. (b) Oriented Line Graph obtained from the graph depicted in panel (a).}
     \label{fig:app:example}
 \end{figure*}

In order to make the above concise description clear, let us follow a very simple example of a graph shown in Fig. \ref{fig:app:example}a. In such a case, knowing the adjacency matrix $\mathbf{A}$ where $A_{ij}=1$ if the nodes $i$ and $j$ share a link and $A_{ij}=0$ otherwise, we can write the row-normalised adjacency matrix $\hat{A}_{ij} = A_{ij}/\sum_k A_{kj}$ as 
\begin{equation}
    \hat{\mathbf{A}} = \left(\begin{array}{cccc}
         0 &  \frac{1}{3} & \frac{1}{3} & \frac{1}{3}\\
         \frac{1}{2} & 0 & 0 & \frac{1}{2}\\
         \frac{1}{3} & \frac{1}{3} & 0 & \frac{1}{3}\\
         \frac{1}{2} & 0 & \frac{1}{2} & 0
    \end{array}\right).
\end{equation}

Solving the eigenproblem $\hat{\mathbf{A}}\mathbf{g} = \lambda \mathbf{g}$ one obtains in this case the following eigenvectors
\begin{equation}
    \mathbf{g} = 
    \begin{blockarray}{ccccc}
& {\scriptstyle \omega_1} & {\scriptstyle \omega_2} & {\scriptstyle \omega_3} & {\scriptstyle \omega_4}\\
\begin{block}{c(cccc)}
{\scriptstyle 1~~} & \sqrt{\frac{2}{13}}  & \frac{\sqrt{2}}{2}  & 0 & -\frac{1}{2}\\
{\scriptstyle 2~~} & -\frac{3}{\sqrt{26}}  & 0 & -\frac{\sqrt{2}}{2} & -\frac{1}{2}\\
{\scriptstyle 3~~} & \sqrt{\frac{2}{13}} & -\frac{\sqrt{2}}{2}  & 0 & -\frac{1}{2}\\
{\scriptstyle 4~~} & -\frac{3}{\sqrt{26}}  & 0  & \frac{\sqrt{2}}{2} & -\frac{1}{2}\\
\end{block}
\end{blockarray}
\end{equation}
and the related eigenvalues $\mathbf{\lambda} = \left(-\frac{2}{3}, -\frac{1}{3}, 0, 1\right)$. Each column of $\mathbf{g}$ corresponds to a different $\omega$ and the consecutive rows are the node numbers. As mentioned before, $\lambda=1$ is not taken into account in further calculations, so $\mathbf{\omega} = \{\omega_1, \omega_2, \omega_3\} = \{\arccos(-\frac{2}{3}), \arccos(-\frac{1}{3}), \frac{\pi}{2}\}$. Now, having calculated $\mathbf{g}$ and $\mathbf{\omega}$ we are able to obtain $C(e,\omega$ and $B(e,\omega)$ as described in Eq. (\ref{eq:cb}). To simplify the outcome we show it as matrices with rows denoted by the graph edges and columns --- by $\mathbf{\omega}$ values:
\begin{widetext}
\begin{equation}
\begin{array}{cc}
    \mathbf{C} = 
    \begin{blockarray}{cccc}
& {\scriptstyle \omega_1} & {\scriptstyle \omega_2} & {\scriptstyle \omega_3}\\
\begin{block}{c(ccc)}
{\scriptstyle e_{12}~~} & -\frac{3 \sqrt{26}}{26} & -\frac{3}{4} & \frac{\sqrt{2}}{2} \\
{\scriptstyle e_{13}~~} & -\frac{1}{13} \sqrt{6} \sqrt{26} & -\frac{\sqrt{3}}{2} & 0 \\
{\scriptstyle e_{14}~~} & -\frac{3 \sqrt{26}}{26} & -\frac{3}{4} & \frac{\sqrt{2}}{2} \\
{\scriptstyle e_{21}~~} & \frac{3 \sqrt{26}}{26} & \frac{3}{4} & -\frac{\sqrt{2}}{2} \\
{\scriptstyle e_{23}~~} & \frac{3 \sqrt{26}}{26} & \frac{3}{4} & -\frac{\sqrt{2}}{2} \\
{\scriptstyle e_{31}~~} & -\frac{1}{13} \sqrt{6} \sqrt{26} & \frac{\sqrt{3}}{2} & 0 \\
{\scriptstyle e_{32}~~} & -\frac{3 \sqrt{26}}{26} & \frac{3}{4} & \frac{\sqrt{2}}{2} \\
{\scriptstyle e_{34}~~} & -\frac{3 \sqrt{26}}{26} & \frac{3}{4} & \frac{\sqrt{2}}{2} \\
{\scriptstyle e_{41}~~} & \frac{3 \sqrt{26}}{26} & \frac{3}{4} & \frac{\sqrt{2}}{2} \\
{\scriptstyle e_{43}~~} & \frac{3 \sqrt{26}}{26} & \frac{3}{4} & \frac{\sqrt{2}}{2}\\
\end{block}
\end{blockarray} &
~~~~~~~~~~~~~~    \mathbf{B} = 
    \begin{blockarray}{cccc}
& {\scriptstyle \omega_1} & {\scriptstyle \omega_2} & {\scriptstyle \omega_3}\\
\begin{block}{c(ccc)}
{\scriptstyle e_{12}~~} & \arctan\frac{\sqrt{5}}{2} & -\arctan\frac{\sqrt{2}}{4} & \frac{\pi }{2} \\
{\scriptstyle e_{13}~~} & -\arctan\sqrt{5} & \arctan\frac{\sqrt{2}}{2} & 0 \\
{\scriptstyle e_{14}~~} & \arctan\frac{\sqrt{5}}{2} & -\arctan\frac{\sqrt{2}}{4} & -\frac{\pi }{2} \\
{\scriptstyle e_{21}~~} & 0 & -\frac{\pi }{2} & 0 \\
{\scriptstyle e_{23}~~} & 0 & \frac{\pi }{2} & 0 \\
{\scriptstyle e_{31}~~} & -\arctan\sqrt{5} & \arctan\frac{\sqrt{2}}{2} & 0 \\
{\scriptstyle e_{32}~~} & \arctan\frac{\sqrt{5}}{2} & -\arctan\frac{\sqrt{2}}{4} & \frac{\pi }{2} \\
{\scriptstyle e_{34}~~} & \arctan\frac{\sqrt{5}}{2} & -\arctan\frac{\sqrt{2}}{4} & -\frac{\pi }{2} \\
{\scriptstyle e_{41}~~} & 0 & -\frac{\pi }{2} & 0 \\
{\scriptstyle e_{43}~~} & 0 & \frac{\pi }{2} & 0\\
\end{block}
\end{blockarray}
\end{array}
\end{equation}
\end{widetext}
Each column of the matrix $\mathbf{C}$ needs to by divided by a corresponding value of $\rho(\omega)$ given by Eq. (\ref{eq:rho}), i.e., in the case of the exemplary graph $\mathbb{\rho} = \{\frac{15}{13},\frac{3}{2},1\}$. In this way we possess the full information needed to evaluate values of $u_1$.

Figure \ref{fig:app:example}b presents an Oriented Line Graph obtained from the graph shown in Fig. \ref{fig:app:example}a, its adjacency matrix $\mathbf{A}_{olg}$ being simply
\begin{equation}
    \mathbf{A}_{olg} =     \begin{blockarray}{ccccccccccc}
& {\scriptstyle e_{12}} & {\scriptstyle e_{13}} & {\scriptstyle e_{14}}\ & {\scriptstyle e_{21}} & {\scriptstyle e_{23}} & {\scriptstyle e_{31}} & {\scriptstyle e_{32}} & {\scriptstyle e_{34}} & {\scriptstyle e_{41}} & {\scriptstyle e_{43}}\\
\begin{block}{c(cccccccccc)}
 {\scriptstyle e_{12}~~} & 0 & 0 & 0 & 0 & 1 & 0 & 0 & 0 & 0 & 0 \\
 {\scriptstyle e_{13}~~} & 0 & 0 & 0 & 0 & 0 & 0 & 1 & 1 & 0 & 0 \\
 {\scriptstyle e_{14}~~} & 0 & 0 & 0 & 0 & 0 & 0 & 0 & 0 & 0 & 1 \\
 {\scriptstyle e_{21}~~} & 0 & 1 & 1 & 0 & 0 & 0 & 0 & 0 & 0 & 0 \\
 {\scriptstyle e_{23}~~} & 0 & 0 & 0 & 0 & 0 & 1 & 0 & 1 & 0 & 0 \\
 {\scriptstyle e_{31}~~} & 1 & 0 & 1 & 0 & 0 & 0 & 0 & 0 & 0 & 0 \\
 {\scriptstyle e_{32}~~} & 0 & 0 & 0 & 1 & 0 & 0 & 0 & 0 & 0 & 0 \\
 {\scriptstyle e_{34}~~} & 0 & 0 & 0 & 0 & 0 & 0 & 0 & 0 & 1 & 0 \\
 {\scriptstyle e_{41}~~} & 1 & 1 & 0 & 0 & 0 & 0 & 0 & 0 & 0 & 0 \\
 {\scriptstyle e_{43}~~} & 0 & 0 & 0 & 0 & 0 & 1 & 1 & 0 & 0 & 0 \\
\end{block}
\end{blockarray}.
\end{equation}
We deliberately refrain from showing the full matrix $\mathbf{g}_{olg}$ of eigenvectors of $\mathbf{A}_{olg}$ as in our case $|E|-|V|=1$ so there is exactly one eigenvector corresponding to $\lambda=-1$ namely

\begin{figure*}[tb]
     \centering
     \includegraphics[width=.7\textwidth]{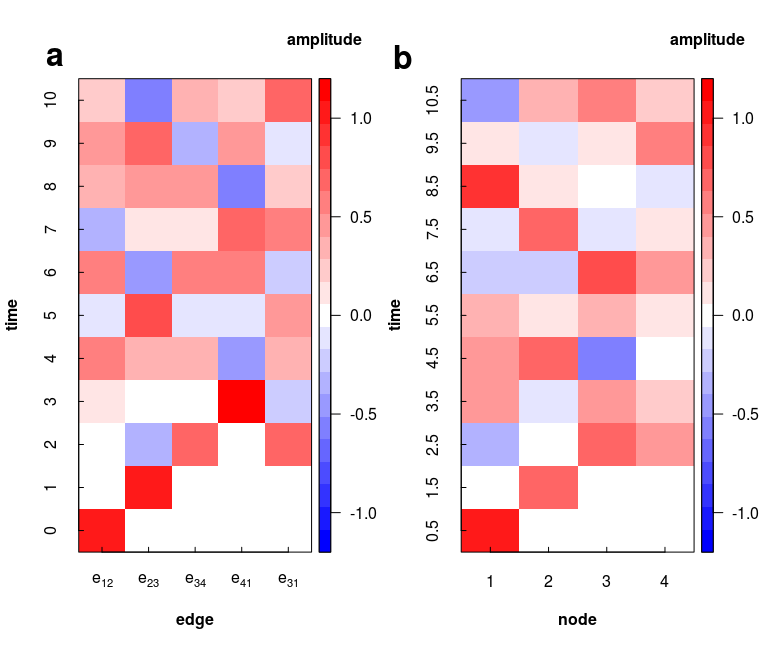}
     \caption{Wave propagation on a graph shown in Fig. \ref{fig:app:example}a for the first 10 time steps (initial condition: a Gaussian wave fully contained on the edge $f=\{1,2\}$ moving towards the vertex $v=2$): (a) the amplitudes on the edges, (b) the amplitudes on the nodes.}
     \label{fig:app:prop}
 \end{figure*}

%\begin{equation}
%\mathbf{C}_{\pi} = \begin{blockarray}{cc}
%& {\scriptstyle 1}\\
%\begin{block}{c(c)}
%{\scriptstyle e_{12}} & \frac{\sqrt{2}}{4}\\
%{\scriptstyle e_{13}} & 0\\
%{\scriptstyle e_{14}} & -\frac{\sqrt{2}}{4}\\
%{\scriptstyle e_{21}} & \frac{\sqrt{2}}{4}\\
%{\scriptstyle e_{23}} & -\frac{\sqrt{2}}{4}\\
%{\scriptstyle e_{31}} & 0\\
%{\scriptstyle e_{32}} & -\frac{\sqrt{2}}{4}\\
%{\scriptstyle e_{34}} & \frac{\sqrt{2}}{4}\\
%{\scriptstyle e_{41}} & -\frac{\sqrt{2}}{4}\\
%{\scriptstyle e_{43}} & \frac{\sqrt{2}}{4}\\
%\end{block}
%\end{blockarray}.
%\end{equation}

\begin{equation}
\mathbf{C}_{\pi} = \begin{blockarray}{cccccccccc}
{\scriptstyle e_{12}} & {\scriptstyle e_{13}} & {\scriptstyle e_{14}}\ & {\scriptstyle e_{21}} & {\scriptstyle e_{23}} & {\scriptstyle e_{31}} & {\scriptstyle e_{32}} & {\scriptstyle e_{34}} & {\scriptstyle e_{41}} & {\scriptstyle e_{43}}\\
\begin{block}{(cccccccccc)}
\frac{\sqrt{2}}{4} & 0 & -\frac{\sqrt{2}}{4} & \frac{\sqrt{2}}{4} & -\frac{\sqrt{2}}{4} & 0 & -\frac{\sqrt{2}}{4} & \frac{\sqrt{2}}{4} & -\frac{\sqrt{2}}{4} & \frac{\sqrt{2}}{4}\\
\end{block}
\end{blockarray}.
\end{equation}
It is now easy to check that if we substitute Eq. (\ref{eq:u}) with the calculated matrices $\mathbf{C}$, $\mathbf{B}$, $\mathbf{\omega}$ and $\mathbf{C}_{\pi}$, and assume that the wave is initially localised on the edge $f=\{1,2\}$, and $t=0$, the amplitude $u=1$ for $e=f=\{1,2\}$ and $u=0$ in any other case, as expected. The first 10 steps of the propagation can be depicted in Fig. \ref{fig:app:prop} (the wave moves towards $v=2$), showing both the amplitudes on the edges (panel a) as well as the nodes (panel b).

\section{Principal component analysis}\label{app:b}
The principal components are a sequence of projections of the set of data in $\mathbb{R}^p$, mutually uncorrelated and ordered in variance in $\mathbb{R}^q$ where $q \leq p$ \cite{hastie2009elements}.  
In other words we transform the feature space such that it becomes orthogonal and each consecutive feature is aligned in the direction maximising the variance of the data and has more variance than the last. We do that by minimising the reconstruction error, i.e., solving:
\begin{equation}
    \min_{\mathbf{V}_q} \sum_{i=1}^N ||(x_i - \bar{x}) - \mathbf{V}_q \mathbf{V}_q^T (x_i - \bar{x})||^2
\end{equation}
where $\mathbf{V}_q$ is a $p \times q$ matrix with $q$ orthogonal unit vectors as columns.
A $p \times p$ matrix $V_q V_q^T$ is the \textit{transformation matrix} that maps each $p$-dimensional observation into its $q$-rank reconstruction. In our case specifically $p=q=100$ and the examples of transformation matrices are represented as heat-maps in Fig.~\ref{fig:wps-ba} and ~\ref{fig:wps-er}.
In general the PCA is known to be a quick and easy method to (i) perform dimensional reduction, (ii) help to visualise high-dimensional data and (iii) aggregate high-dimensional data into a possibly single measure (see, e.g., \cite{gajewski2016key, sienkiewicz2018categorical, choloniewski_calibrated_2020}).
 
\section{K-nearest neighbours}\label{app:c}
In the K-NN classification method the class estimation $\hat{y}(x)$ of a given sample $x$ is taken as a majority vote amongst the members of $N_K(x)$ - the neighbourhood of $x$ defined as $K$ points closest to $x$ \cite{cover1967nearest, hastie2009elements}. 
To determine which points are closest a metric must be chosen and for the purposes of this paper a Euclidean distance was used.
In our case specifically, each observation is a graph represented by its WPS, i.e., each graph is a point in a 100-dimensional space.

\section{Fourier analysis}\label{app:d}
Fourier analysis allows us to convert a given time dependent signal $f(t)$ onto a frequency domain, into $\hat{f}(\omega)$, via a Fourier transform and thus acquire the frequency distribution of said signal as it becomes a linear combination of the trigonometric functions, each corresponding to a particular frequency. A discrete Fourier transform is, as name suggests, a discrete version where the integration is replaced by a summation \cite{arfken1999mathematical}.
Therefore, we consider a problem where one wants to express $f(t)$ as a complex Fourier series:
\begin{equation}
    \hat{f}(\omega) = \sum_{k=0}^{N-1} f(k)e^{2\pi i \omega / N}
\end{equation}
This procedure, as it stands, would require $N^2$ operations (where each operation is a complex multiplication followed by a complex addition), however, Cooley and Tukey in \cite{cooley1965algorithm} presented a method known as the \textit{fast Fourier transform} that allows us to do it in less than $2N \log_2 N$.

\section{Real-world networks}\label{app:e}
We use three real-world networks to test the Fourier transform approach.

Vickers \textit{et al.} \cite{vickers1981representing} collected data from 29 7th grade students from Victoria, Australia.
Students were asked to nominate classmates in several categories, three of which were used to construct this 3-layer network. These three categories were determined by questions -  Who do you get on with in the class? Who are your best friends in the class? Who would you prefer to work with? The graph has 29 nodes and 740 edges in total.

Krackhardt \cite{krackhardt1987cognitive} took a record of relationship between managers in a high-tech company. The graph has 21 nodes and 312 edges in a 3-layer form. Each layer represents a relationship (advice, friendship, "reports to").

Chen \textit{et al.} \cite{chen2006wiring} presented a \textit{Caenorhabditis elegans} multiplex connectome network with 3 layers, 279 nodes and 5863 edges. Each layer corresponds to a different synaptic junction: electric, chemical monadic, and polyadic.
\begin{table*}[!ht]
\begin{tabular}{c|c|c|c|c}
edge $e$ & $C(e,\omega_{N-2})$ & $B(e,\omega_{N-2})$ & $C(e,\omega_{N-1})$ & $B(e,\omega_{N-1})$\\
\hline\hline
$e_{12}$ & $\sqrt{\frac{N-1}{N}}$ & $-\frac{\pi}{2}$ & $\sqrt{\frac{N-1}{N}}$ & 0\\\hline
$e_{1k}$ & $\sqrt{\frac{N-1}{N}}\frac{1}{N-2}$ & $\frac{\pi}{2}$ & $\sqrt{\frac{N-1}{N}}$ & 0\\\hline
$e_{21}$ & $\sqrt{\frac{N-1}{N}}$ & $-\arctan\left( \frac{1}{\sqrt{N(N-2)}}\right)$ & -$\sqrt{\frac{N-1}{N}}$ & $\arctan\left( \sqrt{N(N-2)}\right)$\\
\end{tabular}
\caption{Coefficients $C$ and $B$ for the full graph case; in the third row $k=2,...,N$.}
\label{tab:cb}
\end{table*}

\section{Peak detection using a wavelet transform}\label{app:f}
A wavelet transform is an analogous procedure to the Fourier transform in the sense that we represent a given signal as an orthonormal series \cite{daubechies1992ten}. In case of Fourier those are sine and cosine, while in the wavelet those are the eponymous wavelets. A wavelet is a particularly chosen function that is localised, i.e., it has a finite width, and its family can compose an orthonormal basis for the signal - $s(t)$. 
\begin{equation}
    C(a, b) = \int_{\mathcal{R}} s(t) \frac{1}{\sqrt{a}}\psi\bigg(\frac{t-b}{a}\bigg)dt,~a \in \mathcal{R}^{+}, b \in \mathcal{R}
\end{equation}
In our case the wavelet - $\psi$ - was a Morlet (also known as the Mexican hat) one as per the procedure described in \cite{peakdet} which (simplified) is as follows: perform a continuous wavelet transform (CWT) on the signal, identify the ridge lines by linking local maxima of CWT at each scale level, identify the peaks based on the ridge lines with three rules (quoted verbatim): ``(1) The scale corresponding to the maximum amplitude on the ridge line,which is proportional to the width of the peak, should be within a certain range;(2) The SNR should be larger than a certain threshold;(3) The length of ridge lines should be larger than a certain threshold;''. Here SNR is a \textit{signal to noise ratio}.
 
\section{Exact analytical solutions of wave amplitudes in monoplex and duplex full graphs}\label{app:full}

\subsection{Monoplex full graph}

Here we will consider a wave propagation on the nodes of a full graph (monoplex) topology, i.e., each node is connected to any other in the network, therefore forming a clique of $N$ nodes. In such a case the adjacency matrix $\mathbf{A}$ is a constant matrix filled with ones except for the diagonal which is filled with zeroes.

It is well known that the spectrum of the full graph consists of $N-1$ with multiplicity 1 and  $-1$ with multiplicity $N-1$. As we consider the row-normalised matrix which can be characterised as $\hat{\mathbf{A}}=\frac{1}{N-1}\mathbf{A}$ we have

\begin{equation}
\bm{\lambda} = \{\lambda_1,...,\lambda_{N-1},\lambda_N\} = \left\{-\frac{1}{N-1},...,-\frac{1}{N-1},1\right\} 
\end{equation}
and, consequently, $\omega_i = \arccos\left(-\frac{1}{N-1}\right)$ for $i=1,...,N-1$. As the full graph is $(N-1)$-regular then $\mathbf{e}$, i.e., all-ones vector, is an eigenvector of $\mathbf{A}$ corresponding to $\lambda_N$, while the other vectors can be written as $\mathbf{e}_i - \mathbf{e}_j$ for $i \neq j$, where $\mathbf{e}_i$ is the vector with 1 in position $i$ and 0 elsewhere, e.g., 

\begin{equation}
\mathbf{g} = \left(
\begin{array}{cccccccc}
-1 & -1 & -1 & ... & - 1 & -1 & 1\\
0 & 0 & 0 & ... & 1 & 0 & 1\\
\vdots & \vdots & \vdots & \reflectbox{$\ddots$} & \vdots & \vdots & 1\\
0 & 0 & 1 &  ... & 0 & 0 & 1\\
0 & 1 & 0 & ... & 0 & 0 & 1\\
1 & 0 & 0 & ... & 0 & 0 & 1\\
\end{array}
\right)
\end{equation}

After orthnormalising $\mathbf{g}$ with the Gram-Schmidt procedure and renumbering the matrix so that the first row becomes the last one we obtain the following $N-1 \times N$ eigenvector matrix $\mathbf{g}_M$:

\begin{widetext}
\begin{equation}
\mathbf{g_M} = \left(
\begin{array}{cccccccc}
 0 & 0 & 0 & 0 & 0 & 0 & 0 & \sqrt{\frac{N-1}{N}} \\
 
 \vdots & \vdots & \vdots & \vdots & \vdots & 0 & \sqrt{\frac{N-2}{N-1}} & -\frac{1}{\sqrt{(N-1)N}}\\
 \vdots & \vdots & \vdots & \vdots & 0 & \sqrt{\frac{j}{j+1}} & \cdots &  \vdots\\
 \vdots & \vdots & \vdots & 0 & \reflectbox{$\ddots$} & -\frac{1}{\sqrt{j(j+1)}} & \cdots & \vdots\\
 
 \vdots & \vdots & 0 & \reflectbox{$\ddots$} & \cdots & \vdots & \cdots & \vdots\\
 
 \vdots & 0 & \sqrt{\frac{3}{4}} & \cdots & ... & -\frac{1}{\sqrt{j(j+1)}} & \cdots & -\frac{1}{\sqrt{(N-1)N}}\\

 0 & \sqrt{\frac{2}{3}} & -\frac{1}{\sqrt{3\cdot4}} & \cdots & ... & -\frac{1}{\sqrt{j(j+1)}} & \cdots & -\frac{1}{\sqrt{(N-1)N}}\\

 \sqrt{\frac{1}{2}} & -\frac{1}{\sqrt{2\cdot3}} & -\frac{1}{\sqrt{3\cdot4}} & \cdots & ... & -\frac{1}{\sqrt{j(j+1)}} & \cdots & -\frac{1}{\sqrt{(N-1)N}}\\

 -\frac{1}{\sqrt{1 \cdot 2}} & -\frac{1}{\sqrt{2\cdot3}} & -\frac{1}{\sqrt{3\cdot4}} & \cdots & ... & -\frac{1}{\sqrt{j(j+1)}} & \cdots & -\frac{1}{\sqrt{(N-1)N}}\\
\end{array}
\right)
\label{eq:gm}
\end{equation}
\end{widetext}

Note that we deliberately omit the eigenvector corresponding to $\omega_N$ which consists of $N$ values $1/\sqrt{N}$ gathered in the column vector  
\begin{equation}
\mathbf{C}_M = \left(\frac{1}{\sqrt{N}},...,\frac{1}{\sqrt{N}}\right)^{T}    
\label{eq:cm}
\end{equation}
as it does not play any role in further calculations.

Let us now assume, that the wave is initially contained on edge $f=e_{uv}$. We are bound to obtain two expressions: one showing the amplitude of the wave at the node $v$ which can be simply calculated as $u_n(f,f,t)$ and the other one, showing the sum of the amplitudes for all nodes which can be expressed as 
\begin{equation}
u_{all} = u_n(e_{vu},f,t) + \sum_{i=1,..,N,i \neq u}u_n(e_{ui},f,t).
\label{eq:all}
\end{equation}

We will further assume, without losing any generality, that $f=e_{12}$. Such a setting greatly simplifies further calculations of $u_n$ as in this case the only non-zero inputs come from eigenvalues $\omega_{N-2}$ and $\omega_{N-1}$. For any other $\omega_i$ with $i=1,...,N-3$ we have $C(f,\omega_i)=0$. Respective coefficients $C$ and $B$ for $\omega_{N-2}$ and $\omega_{N-1}$ are gathered in Table \ref{tab:cb}  

As mentioned in the Appendix \ref{app:a}, in order to have a proper form of the eigenfunction $\phi$ one needs to calculate the normalisation factor $\rho(\omega)$. Next we show how to obtain $\rho(\omega_{N-1})$: Eq. (\ref{eq:rho}) states that it is necessary to sum of over all the edges in the networks, however, in the case of $\omega_{N-1}$ there are only two different terms. The first one $\rho_1$ comes from $(N-1)$ pairs $e_{12},e_{13},...,e_{1N}$, the second one $\rho_2$ from all the other $\frac{1}{2}(N-1)(N-2)$ pairs, i.e., $e_{23},...,e_{2N},e_{34},...,e_{N-1N}$ so 

\begin{equation}
    \rho(\omega_{N-1}) = \sqrt{(N-1)\rho_1+\frac{1}{2}(N-1)(N-2)\rho_2}
\label{eq:rhom}
\end{equation}

The values of $C$ and $B$ for $\rho_1$ are shown in Table (\ref{tab:cb}) -- taking into account that $B=0$, the value of $\rho_1$ simplifies to:

\begin{equation}
    \rho_1 = \frac{N-1}{N} \left[\frac{1}{2} + \frac{\sin \left[2 \arccos \left(-\frac{1}{N-1}\right)\right]}{4 \arccos \left(-\frac{1}{N-1}\right)}\right].
\end{equation}
Substituting $\sin(2 \arccos x)$ with $2x\sqrt{1-x^2}$ we arrive at
\begin{equation}
    \rho_1 = \frac{N-1}{N} \left[\frac{1}{2} - \frac{2 \frac{\sqrt{N(N-2)}}{(N-1)^2}}{4 \arccos \left(-\frac{1}{N-1}\right)}\right].
\label{eq:rho1}
\end{equation}

In the case of $\rho_2$ the values $C$ and $B$ can be written as
\begin{align}
C(e_{ik},\omega_{N-1}) &= -\sqrt{\frac{2}{N(N-2)}}\\
B(e_{ik},\omega_{N-1}) &= -\arctan\sqrt{\frac{N}{N-2}}
\end{align}
for $i=2,...,N-1$ and $k=i+1,...,N$ which results in the following form of $\rho_2$:

\begin{widetext}
\begin{equation}
    \rho_2 = \frac{2}{N(N-2)} \left[\frac{1}{2} + \frac{\sin \left[2 \arccos \left(-\frac{1}{N-1}\right) - 2\arctan\sqrt{\frac{N}{N-2}}\right] + \sin\left(2\arctan\sqrt{\frac{N}{N-2}}\right)}{4 \arccos\left(-\frac{1}{N-1}\right)}\right].
\end{equation}
\end{widetext}

Here, making use of the fact that $\arccos x = 2 \arctan \frac{\sqrt{1-x^2}}{1+x}$ we arrive at
\begin{equation}
    \rho_2 = \frac{2}{N(N-2)} \left[\frac{1}{2} + \frac{2 \frac{\sqrt{N(N-2)}}{N-1}}{4 \arccos \left(-\frac{1}{N-1}\right)}\right].
\label{eq:rho2}
\end{equation}
After substituting Eq. (\ref{eq:rhom}) with $\rho_1$ and $\rho_2$ given by Eqs. (\ref{eq:rho1}) and (\ref{eq:rho2}), and performing some short algebra we arrive simply at
\begin{equation}
    \rho(\omega_{N-1}) = \sqrt{\frac{N-1}{2}}.
\label{eq:rhof}
\end{equation}
With similar calculations it possible to show that $\rho(\omega_{N-2}) = \rho(\omega_{N-1})$ (in fact, this the case for any $\omega_i$ other than $\omega_N$).

\begin{table*}[!ht]
\begin{tabular}{c|c|c|c|c}
edge $e$ & $C(e,\omega_{N-2})$ & $B(e,\omega_{N-2})$ & $C(e,\omega_{N-1})$ & $B(e,\omega_{N-1})$\\
\hline\hline
$e_{12}$ & $\frac{N}{\sqrt{2(N-1)(N+2)}}$ & $\frac{\pi}{2}$ & $-\sqrt{\frac{N^2-2}{2(N-1)(N+2)}}$ & $-\arctan\left(\frac{1}{N-1}\sqrt{\frac{N-2}{N+2}}\right)$\\\hline
$e_{1k}$ & $\frac{N}{N-2}\frac{1}{\sqrt{2(N-1)(N+2)}}$ & $-\frac{\pi}{2}$ & $-\sqrt{\frac{N^2-2}{2(N-1)(N+2)}}$ & $-\arctan\left(\frac{1}{N-1}\sqrt{\frac{N-2}{N+2}}\right)$\\\hline
$e_{21}$ & $-\frac{1}{\sqrt{2(N-1)(N+2)}}$ & $-\arctan\left(\frac{2}{\sqrt{N^2-4}}\right)$ & $\sqrt{\frac{N^2-2}{2(N-1)(N+2)}}$ & $\arctan\left((N+1)\sqrt{\frac{N-2}{N+2}}\right)$\\\hline
\end{tabular}
\caption{Coefficients $C$ and $B$ for the duplex full graph case; in the third row $k=2,...,N$.}
\label{tab:cbdup1}
\end{table*}

\begin{table*}[!ht]
\begin{tabular}{c|c|c|c|c|c|c}
edge $e$ & $C(e,\omega_{2N-3})$ & $B(e,\omega_{2N-3})$ & $C(e,\omega_{2N-2})$ & $B(e,\omega_{2N-2})$ & $C(e,\omega_{2N-1})$ & $B(e,\omega_{2N-1})$\\
\hline\hline
$e_{12}$ & $\sqrt{\frac{N-2}{2(N-1)}}$ & $-\frac{\pi}{2}$ & 
$\sqrt{\frac{(N-1)^2+1}{2N(N-1)}}$ & $\arctan\left(\frac{1}{N-1}\right)$ & $-\sqrt{\frac{1}{2(N-1)}}$ & $-\arctan\left(\sqrt{\frac{1}{(N-1)}}\right)$\\\hline
$e_{1k}$ & $\sqrt{\frac{1}{2(N-2)(N+1)}}$ & $\frac{\pi}{2}$ & 
$\sqrt{\frac{(N-1)^2+1}{2N(N-1)}}$ & $\arctan\left(\frac{1}{N-1}\right)$ & $-\sqrt{\frac{1}{2(N-1)}}$ & $-\arctan\left(\sqrt{\frac{1}{(N-1)}}\right)$\\\hline
$e_{21}$ & $\sqrt{\frac{N-2}{2(N-1)}}$ & $0$ & 
$-\sqrt{\frac{(N-1)^2+1}{2N(N-1)}}$ & $\arctan\left(N-1\right)$ & $-\sqrt{\frac{1}{2(N-1)}}$ & $-\arctan\left(\sqrt{\frac{1}{(N-1)}}\right)$\\\hline\end{tabular}
\caption{Coefficients $C$ and $B$ for the duplex full graph case (cntd).}
\label{tab:cbdup2}
\end{table*}

We can now use the above calculated values of $C, B$ and $\rho$ to express the wave amplitude at the nodes of the full graph by the means of Eq. (\ref{eq:un}) [note that in order to simplify the equation we use the transformation of time $t \rightarrow t + \frac{1}{2}$, thus $t=0$ means that the wave has arrived at the first node]. We shall start with $u_n(e_{12},e_{12},t)$ as the simplest case:
\begin{widetext}
\begin{equation}
\begin{split}
u_n(e_{12},e_{12},t) &= \sum\limits_{i=\{N-2,N-1\}}\frac{C^2(e_{12},\omega_{i})}{2\rho^2(\omega_i)}\left[\cos(\omega_i t) + \cos\right((t+2)\omega+2B(e_{12},\omega_{i})\left)\right]+\frac{1}{E}\\
&=\frac{1}{2\rho^2(\omega_{N-2})}\frac{N-1}{N}\left[\cos t \omega_{N-2} + \cos[ (t+2) \omega_{N-2} - \pi]\right]+\\
&+\frac{1}{2\rho^2(\omega_{N-1})}\frac{N-1}{N}\left[\cos t \omega_{N-1} + \cos[ (t+2) \omega_{N-1}]\right] + \frac{2}{N(N-1)}\\
\end{split}
\end{equation}
\end{widetext}
Making use of Eq. (\ref{eq:rhof}) and the fact that $\cos(x-\pi)=-\cos(x)$ as well as denoting $\omega_{N-2}=\omega_{N-1}=\omega$ we arrive simply at
\begin{equation}
u_n(e_{12},e_{12},t) = \frac{2}{N} \cos(t \omega) + \frac{2}{N(N-1)} 
\label{eq:u12}
\end{equation}
with $\omega = \arccos \left(-\frac{1}{N-1}\right)$. In the same manner one can show that
\begin{widetext}
\begin{equation}
u_n(e_{1k},e_{12},t) = \frac{1}{N(N-2)}\left((N-3)\cos(t \omega) + (N-1)\cos((t+2) \omega) \right)+ \frac{2}{N(N-1)}
\label{eq:u1k}
\end{equation}
\end{widetext}

for $k=3,...,N$ and

\begin{widetext}
\begin{equation}
u_n(e_{21},e_{12},t) = -\frac{2}{N(N-1)}\left(\cos(t \omega) + \sqrt{N(N-2)}\sin((t+2) \omega) \right)+ \frac{2}{N(N-1)}
\label{eq:u21}
\end{equation}
\end{widetext}

In this way, Eq. (\ref{eq:u12})-(\ref{eq:u21}) fully determine the wave amplitude at every node in the considered monoplex full graph. They can now be put into Eq. (\ref{eq:all}) to show that
\begin{equation}
    u^{M}_{all}(t) = \frac{2}{N-1}
\end{equation}
which proves that the sum of the node amplitudes is constant for any value of $t$.

\subsection{Duplex full graph}

In this part we consider a full graph duplex, i.e., a network consisting of two cliques (full graphs) of $N$ nodes each connected in such a way that the node $i$ from the first clique links with the node $i + N$ from the second one ($i=1,...,N$). Assuming that $\mathbf{A}_N$ is an $N \times N$ adjacency matrix of a full graph (as in the previous section) and $\mathbf{I}_N$ is an $N \times N$ identity matrix, we can describe the topology of a full graph duplex with its $2N \times 2N$ adjacency matrix $\mathbf{A}_d$

\begin{equation}
\mathbf{A}_d = \left(
\begin{array}{cc}
\mathbf{A}_N & \mathbf{I}_N\\
\mathbf{I}_N & \mathbf{A}_N\\
\end{array}
\right)
\end{equation}

Owing to the symmetry of the system (each node has exactly $N$ neighbours - $N-1$ in the layer it belongs to and one that connects it to the second layer) the row-normalised matrix can be characterised simply as $\hat{\mathbf{A}}_d=\frac{1}{N}\mathbf{A}_d$.

In such a setting we can obtain the following orthonormalised $2N \times 2N$ matrix of the eigenvectors for the duplex network

\begin{equation}
\mathbf{g}_D = \frac{1}{\sqrt{2}}\left(
\begin{array}{cccc}
-\mathbf{g}_M & \mathbf{g}_M & ~~\mathbf{C}_M & \mathbf{C}_M\\
~~\mathbf{g}_M & \mathbf{g}_M & -\mathbf{C}_M & \mathbf{C}_M\\
\end{array}
\right)
\end{equation}
where $\mathbf{g}_M$ and $\mathbf{C}_M$ are the matrices given by Eq. (\ref{eq:gm}) and (\ref{eq:cm}), and the corresponding set of $2N$ eigenvalues is

\begin{equation}
\begin{split}
\bm{\lambda} &= \{\lambda_1,...,\lambda_{N-1},\lambda_N,...,\lambda_{2N-2},\lambda_{2N-1},\lambda_{2N}\} =\\ &=\left\{-\frac{2}{N},...,-\frac{2}{N},0,...,0,\frac{N-2}{N},1\right\} 
\end{split}
\end{equation}
and, consequently,
$\omega_1 = ... = \omega_{N-1} = \arccos\left(-\frac{2}{N}\right)$, $\omega_N = ... = \omega_{2N-2} = \frac{\pi}{2}$, $\omega_{2N-1}=\arccos\left(\frac{N-2}{N}\right)$

As in the previous case we assume that the wave is initially placed on the edge $f=e_{12}$ and our goal is to find the wave amplitudes on each of the nodes $1,...,N$, as well as the total amplitude on a given layer. Again, the choice of the initial edge is a direct consequence of the structure of $\mathbf{g}_D$ as it restricts the set of eigenvalues needed to calculate $u_n$ to $\omega_{N-2}$, $\omega_{N-1}$, $\omega_{2N-3}$, $\omega_{2N-2}$ and $\omega_{2N-1}$ -- all others result in $C(f,\omega)=0$. The values of $C$ and $B$ necessary to compute $u_n(e_{12},e_{12},t)$, $u_n(e_{1k},e_{12},t)$ for $k=2,...,N$ and $u_n(e_{21},e_{12},t)$ obtained from $\mathbf{g}_D$ are summed up in Tables \ref{tab:cbdup1} and \ref{tab:cbdup2}. Following similar calculations as in the monoplex case one can show that the normalisation factor is

\begin{figure*}[!ht]
     \centering
     \includegraphics[width=.8\textwidth]{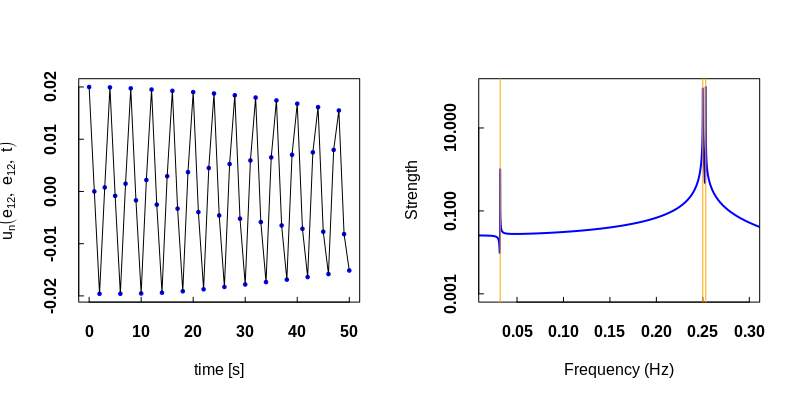}
     \caption{Full graph duplex with $N=100$ nodes in each layer: (left) the amplitude at the node $v=2$ over time; lines are guidance to eye connecting theoretical predictions from Eq. (\ref{eq:undup}) with the coefficients $\mathbf{c}^{12}$ and $\mathbf{s}^{12}$ that are exactly covered with points obtained from numerical simulations; (right) the power spectrum of the signal shown in the left panel [the series consists of 10000 elements]. The vertical lines are drawn for frequencies equal to $\omega_1 / 2\pi$, $\omega_N / 2\pi$ and $\omega_{2N-1} / 2\pi$.}
     \label{fig:undup}
 \end{figure*}

\begin{equation}
\begin{split}
    \rho(\omega_{N-2}) &= \rho(\omega_{N-1}) = \rho(\omega_{2N-3}) =\\ &= \rho(\omega_{N-2}) = \rho(\omega_{2N-1}) = \sqrt{\frac{N}{2}}
\end{split}
\end{equation}

Using the above-calculated values of $C$, $B$ and $\rho$ it is possible to give exact solutions for the amplitude at any node $v$ as  

\begin{equation}
u_n(e_{uv},e_{12},t) = \frac{1}{N^2}\left[1+\sum\limits_i\left(c^{uv}_i\cos \omega_i t+s^{uv}_i\sin \omega_i t \right)\right]
\label{eq:undup}
\end{equation}

where 

\begin{equation}
\begin{split}
\bm{\omega} &= \{\omega_1, \omega_N, \omega_{2N-1}\}\\
&= \left\{\arccos\left(-\frac{2}{N}\right), \frac{\pi}{2}, \arccos\left(\frac{N-2}{N}\right)\right\} 
\end{split}
\end{equation}

and the respective coefficients for $u_n(e_{12},e_{12},t)$ are given by

\begin{align}
\mathbf{c}^{12} &= \{N-1, N-1, 1\}\\
\mathbf{s}^{12} &= \left\{-\sqrt{\frac{N-2}{N+2}}, 1, -\frac{1}{\sqrt{N-1}}\right\}\nonumber
\end{align}

while in the case of $u_n(e_{1k},e_{12},t)$, $k=2,...,N$ they are

\begin{align}
\mathbf{c}^{1k} &= \{-1, -1, 1\}\\
\mathbf{s}^{1k} &= \left\{\sqrt{\frac{N+2}{N-2}}, 1, -\frac{1}{\sqrt{N-1}}\right\}\nonumber
\end{align}

and for $u_n(e_{21},e_{12},t)$, $k=2,...,N$ we have

\begin{align}
\mathbf{c}^{21} &= \{-1, -1, 1\}\\
\mathbf{s}^{21} &= \left\{-(N+1)\sqrt{\frac{N-2}{N+2}},-(N-1),  -\frac{1}{\sqrt{N-1}}\right\}\nonumber
\end{align}
Finally by calculating
\begin{equation}
    u^{D}_{all}(t) = u_n(e_{12},e_{12},t) + (N-2)u_n(e_{1k},e_{12},t) + u_n(e_{21},e_{12},t)
\end{equation}
we arrive at the expression giving the sum of amplitudes over all nodes in a single layer in the duplex full graph which reads
\begin{equation}
    u^{D}_{all}(t) = \frac{1}{N} \left[1 + \cos\left(t \omega_{2N-1}\right) -\frac{1}{\sqrt{N-1}}\sin\left(t \omega_{2N-1}\right) \right].
    \label{eq:ualldup}
\end{equation}

Formula (\ref{eq:undup}) indicates that in the case of a single node all three eigenvalues $\arccos\left(-\frac{2}{N}\right), \frac{\pi}{2}$ and  $\arccos\left(\frac{N-2}{N}\right)$ can be recovered by observing the wave amplitude over time at such a node. Indeed the outcomes of the power spectrum of numerical implementation of the wave propagation algorithm shown in Fig. \ref{fig:undup} confirm this claim. It should be noted, though, that once $N$ is sufficiently large, two eigenvalues $\arccos\left(-\frac{2}{N}\right)$ and $\frac{\pi}{2}$ will tend to merge and $\frac{N-2}{N}$ shall approach 1, giving in result two frequencies: $1/4$ and 0. On the other hand, Eq. (\ref{eq:undup}) clearly shows that if the sum of the amplitudes in a single layer is observed then we are able to recover only one eigenvalue, namely $\omega=\arccos\left(\frac{N-2}{N}\right)$, this fact is depicted in Fig. \ref{fig:ualldup} in the main text.   

\section{Amplitude signal on ER graph}
\label{app:er}
\begin{figure*}[tb]
     \centering
     \includegraphics[width=\textwidth]{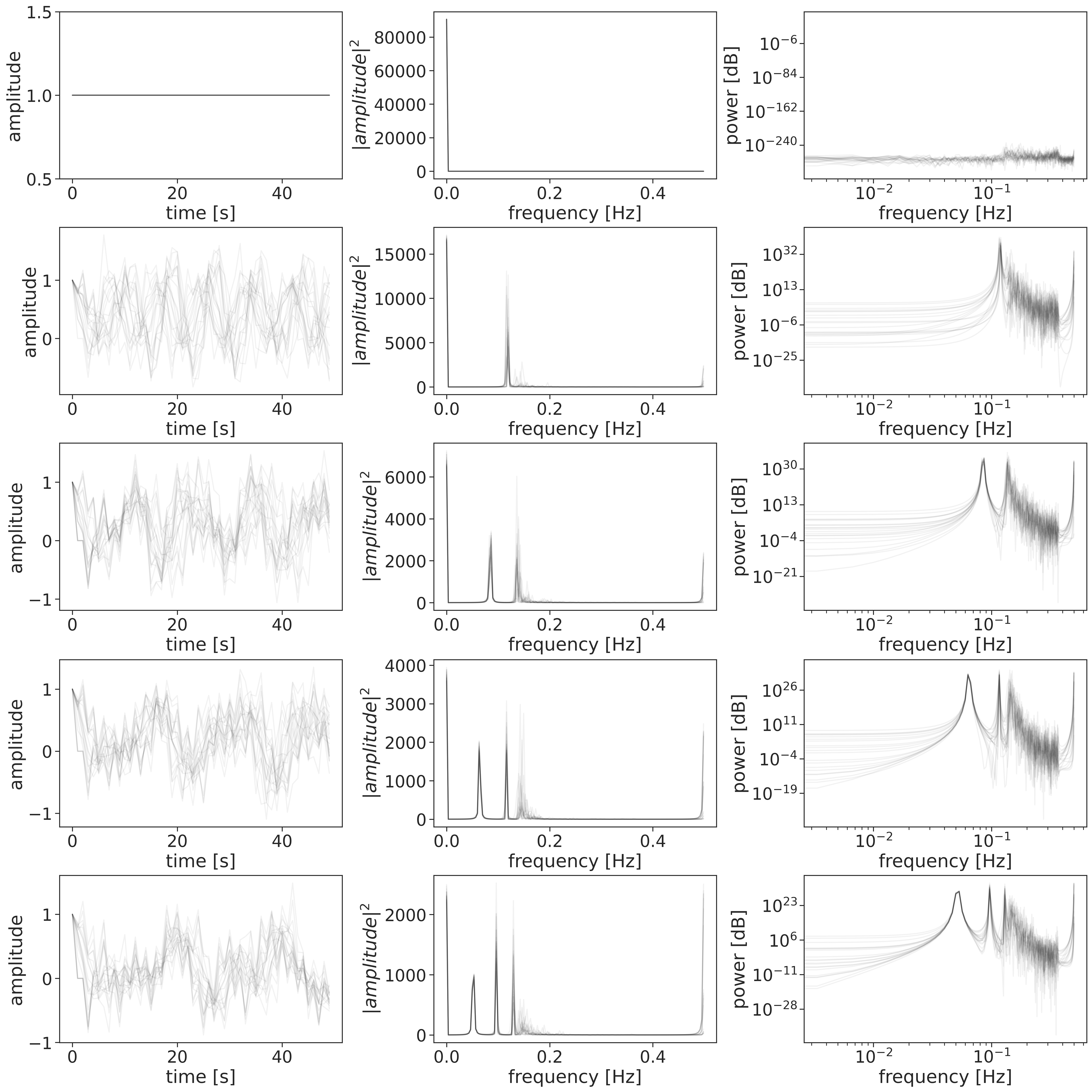}
     \caption{Sum of the amplitudes time evolution as measured on the only visible layer (left). A fast Fourier transform of this signal (centre) and its power spectrum (right). Each row represents a different number of layers (1 to 5 going top to bottom). Simulations were conducted on 20 independent realisations of an ER graph ($N=50$, $\langle k \rangle=6$) per row, overlaid with transparency. Note: for the purposes of the Fourier analysis we use signals of length $10^3$s.}
     \label{fig:fft-er}
 \end{figure*}

As discussed in detail in the main text, it is possible to observe the oscillations in the known layer induced by the existence of other (unknown) layers. The exact nature of these oscillations depends on the underlying structures of the whole graph and while there is some visible variation on the ER graph as compared to the BA model (see Fig. \ref{fig:fft-er} below and \ref{fig:fft-ba} in the main text). This is similar to the case of WPS - see Fig. \ref{fig:wps-ba} and \ref{fig:wps-er} in the main text. Qualitatively these results only further support what can already be seen in Fig. \ref{fig:fft-ba}, i.e., that clear, discernible and consistent peaks in the power spectrum can be detected. 

% \clearpage

% \bibliographystyle{apsrev4-2}  
\bibliography{refs}

\end{document}